\documentclass[letterpaper]{article} % DO NOT CHANGE THIS
\usepackage{aaai23}  % DO NOT CHANGE THIS
\usepackage{times}  % DO NOT CHANGE THIS
\usepackage{helvet}  % DO NOT CHANGE THIS
\usepackage{courier}  % DO NOT CHANGE THIS
\usepackage[hyphens]{url}  % DO NOT CHANGE THIS
\usepackage{graphicx} % DO NOT CHANGE THIS
\usepackage{natbib}  % DO NOT CHANGE THIS AND DO NOT ADD ANY OPTIONS TO IT
\usepackage{caption} % DO NOT CHANGE THIS AND DO NOT ADD ANY OPTIONS TO IT
\usepackage{multirow}
\usepackage{bm, amsmath}
\usepackage{makecell}
\usepackage{rotating}
\usepackage{makecell}
\usepackage{graphicx,subfigure}
\usepackage{booktabs}
\usepackage{rotating}
\usepackage{amssymb}
\usepackage{tablefootnote}
\usepackage{url}

\usepackage[title]{appendix}

\usepackage{xcolor}
\usepackage{colortbl}

\newcommand{\ie}{\emph{i.e., }}
\newcommand{\eg}{\emph{e.g., }}

 \nocopyright % -- Your paper will not be published if you use this command
\title{FakeSV: A Multimodal Benchmark with Rich Social Context \\ for Fake News Detection on Short Video Platforms}

\author{
	Peng Qi\textsuperscript{\rm 1,2}, %$^{1,2}$
	Yuyan Bu\textsuperscript{\rm 1,2},
	Juan Cao\textsuperscript{\rm 1,2}\thanks{Corresponding author.}, 
	Wei Ji\textsuperscript{\rm 3}, 
	Ruihao Shui\textsuperscript{\rm 3}, \\
	Junbin Xiao\textsuperscript{\rm 3}, 
	Danding Wang\textsuperscript{\rm 1},
	Tat-Seng Chua\textsuperscript{\rm 3}
}

\affiliations {
    \textsuperscript{\rm 1} Key Lab of Intelligent Information Processing, Institute of Computing Technology, CAS\\
    \textsuperscript{\rm 2} University of Chinese Academy of Sciences\\
    \textsuperscript{\rm 3} National University of Singapore\\
    \{qipeng,caojuan,wangdanding\}@ict.ac.cn, buyuyan22@mails.ucas.ac.cn,\\
    \{jiwei,dcscts\}@nus.edu.sg, ruihaoshui@u.nus.edu, junbin@comp.nus.edu.sg
}

\usepackage{bibentry}
\begin{document}

\maketitle

\begin{abstract}
Short video platforms have become an important channel for news sharing, but also a new breeding ground for fake news.
To mitigate this problem, research of fake news video detection has recently received a lot of attention. 
Existing works face two roadblocks: the scarcity of comprehensive and large-scale datasets and insufficient utilization of multimodal information. 
Therefore, in this paper, we construct the largest Chinese short video dataset about fake news named FakeSV, which includes news content, user comments, and publisher profiles simultaneously. To understand the characteristics of fake news videos, we conduct exploratory analysis of FakeSV from different perspectives. Moreover, we provide a new multimodal detection model named SV-FEND, which exploits the cross-modal correlations to select the most informative features and utilizes the social context information for detection. Extensive experiments evaluate the superiority of the proposed method and provide detailed comparisons of different methods and modalities for future works.
\end{abstract}

\section{Introduction}
With the prevalence of short video platforms (\eg Tiktok), they have become an important channel for news sharing \cite{case-newssv}.
Besides professional news outlets, ordinary users also upload news videos happening around them. 
 However, this openness have led to such platforms becoming a new breeding ground for fake news.
Video modality is more powerful in spreading fake news than other modalities, and thus will exacerbate the influences of fake news \cite{bg-fakevideo}. 
Therefore, studying video-form fake news is important for detecting and intervening the spread of fake news in a timely manner.

\begin{figure}[tbp]
\setlength{\abovecaptionskip}{0cm}
	\centering
	\subfigure[]
	{
	\includegraphics[width=.21\textwidth, height=.25\textwidth]{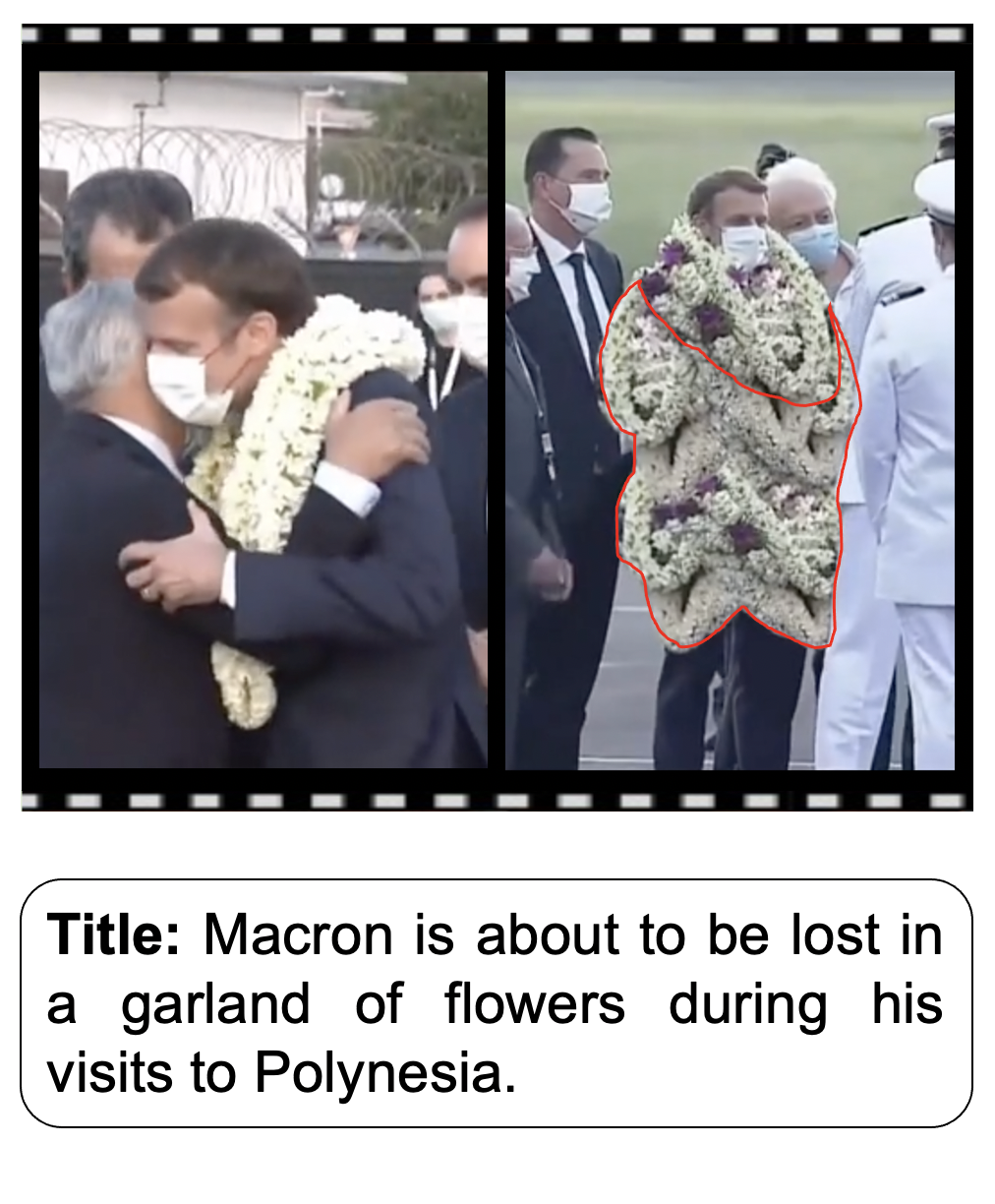}
	\label{fig:case-macron}
	}
	\subfigure[]{
	\includegraphics[width=.22\textwidth, height=.25\textwidth]{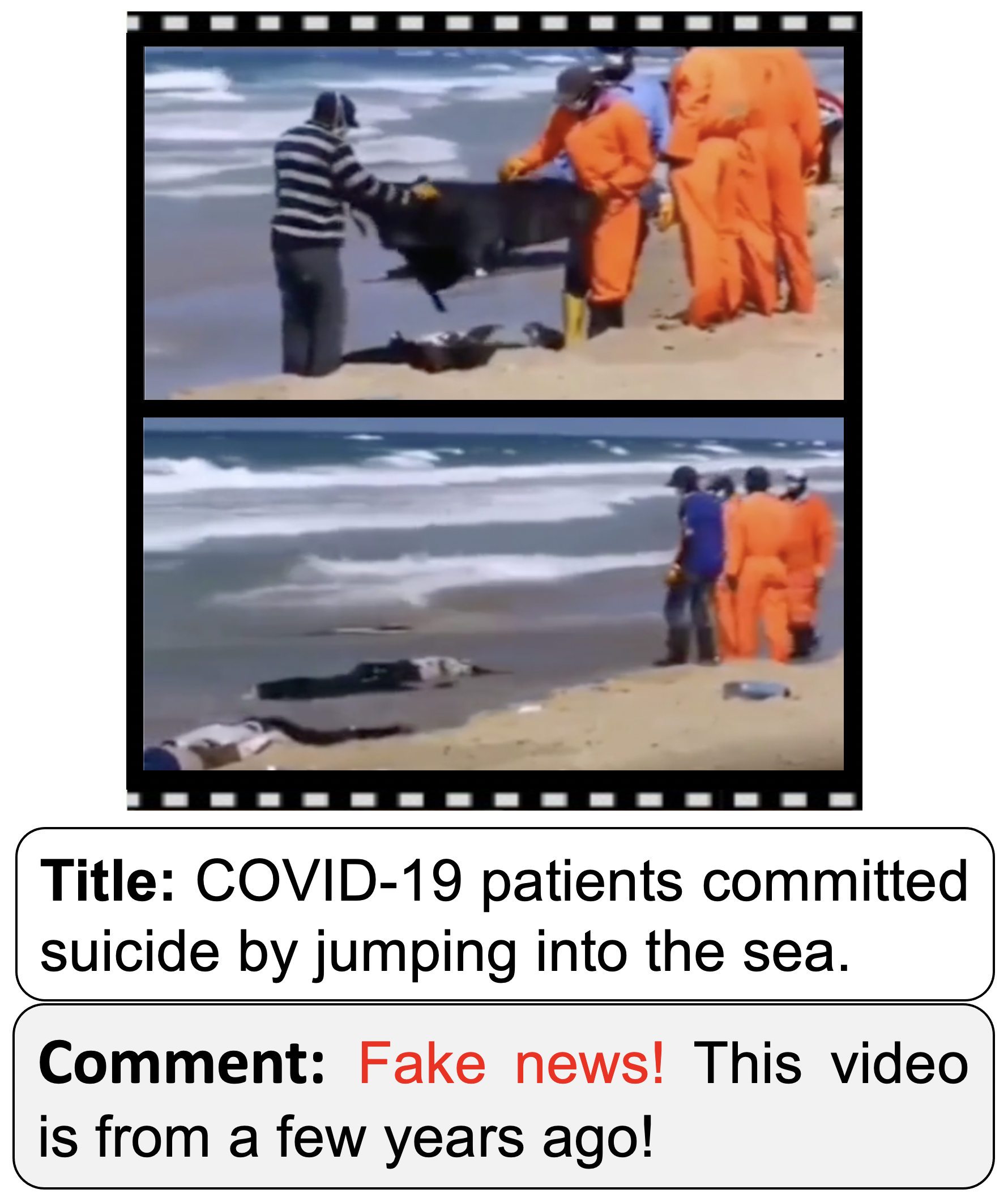}
	\label{fig:case-fox}
	}
	\caption{Typical cases of fake news videos. Figure (a) shows a fake news video of which the content of some frames have been tampered. The spliced region is marked in red line.  
 (b) shows a fake news video that is contextually inappropriate, where the comments provide key clues.  }
	\label{fig:case}
\vspace{-.25em}
\end{figure}

Fake news videos refer to those that the video content and title jointly describe a piece of news that is verifiably false.
Compared with traditional text-based or text-image fake news detection, fake news video detection presents unique challenges: 
First, video-form fake news has more modalities and thus more information. 
We need to select the most informative clues from multiple modalities and fuse heterogeneous information to understand the news content and detect fake news.
Second, different from traditional video platforms such as YouTube, short video platforms provide advanced video-editing functions (\eg text box and virtual background) for users to conveniently modify the videos. 
It weakens the discriminability of the visual content in measuring its truthfulness as both fake and real news videos could be modified and exhibit editing traces.
Moreover, fake news videos typically contain real news videos where only some frames or the accompanying titles have been maliciously modified to alter their meanings (as shown in Figure~\ref{fig:case}).
All these characteristics make it inadequate to study fake news videos from news content only.
Thus, we need to explore auxiliary social context information, such as user comments \cite{defend} (as shown in Figure~\ref{fig:case-fox}) and user profiles \cite{profile}, to help the detection. 

Most works studying fake news focus on text-based or text-image fake news, while few works pay attention to video-form fake news.
There are two major limitations of existing works: 
1) \textbf{Dataset}: 
Although there are several datasets for fake news video detection, the majority of them are constructed on traditional video platforms and contain only a few hundred instances, and none of them provides information on news content, user comments, and user profiles simultaneously.  
Further, the lack of comprehensive datasets with rich modalities hinders the fair comparisons of existing works that use different modalities for detection. 
2) \textbf{Detection methods}: Most works targeting video-based fake news detection are still in the preliminary stage of hand-crafted features. 
Additionally, none of them considers all modalities and thus inevitably misses some important clues. 

To bridge the gap of fake news video datasets, we propose a large-scale Chinese \textbf{Fake} News \textbf{S}hort \textbf{V}ideo dataset (\textbf{FakeSV} for short) that includes both complete news content and rich social context.
The abundant features in this dataset not only provide an opportunity to evaluate different approaches for fake news detection, but also help understand the diffusion of fake news and its intervention.
To study the characteristics of fake news videos, we comprehensively conduct exploratory analysis of FakeSV from different perspectives including the multimodal news content, social context, and propagation, which could shed light on detection strategies.

To tackle the challenges of fake news video detection, we propose a multimodal model named \textbf{SV-FEND} (\textbf{S}hort \textbf{V}ideo \textbf{F}ak\textbf{E} \textbf{N}ews \textbf{D}etection) as a new baseline on FakeSV. 
The proposed model utilizes the co-attention mechanism to enhance the multimodal content representations by spotting the important features, and fuses them with the social context features by the self-attention mechanism.
We conduct extensive experiments comparing the proposed model and existing baseline methods on FakeSV. The results validate the superiority of SV-FEND and provide detailed insights of different methods and modalities for future works.

Our contributions are summarized in three aspects: 
\begin{itemize}
    \item We construct the largest Chinese fake news short video dataset, namely FakeSV.
    This dataset contains complete news contents and rich social context, and thus can support a wide range of research tasks related to fake news. We also provide in-depth statistical analysis.
    \item We provide a new multimodal baseline method SV-FEND that captures the multimodal correlations to enhance the news content representations and utilizes the signals of social context to help the detection. % provide
    \item We conduct extensive experiments with the proposed model and existing SOTA methods on FakeSV, which validate the superiority of the proposed method and provide detailed comparisons of different methods and modalities for future works.
\end{itemize}

\section{Related Work} 

\subsubsection{Datasets.}

As video-form fake news has drawn increasing attention, there have been efforts in constructing datasets to detect fake news videos.
Table~\ref{tab:datasets} summarizes existing datasets in aspects of richness of features, size, domain, language, accessibility, and sources.
Because annotators need to spend more time to watch and understand the video content,  
video-form fake news datasets are typically smaller than text-based and text-image datasets \cite{datasetsurvey}, \eg 5 of 7 datasets only contain less than 1000 instances. Moreover, most of the small-scale datasets provide limited features, only focus on one domain, and do not provide public access to datasets. Thus they are not suitable to study fake news especially in developing detection models that generalize well to new domains. 
In particular, \citet{fvc} build the largest video-form fake news dataset.
They first annotate an initial set of 380 videos and then collect related videos by searching the multi-lingual titles of these seed videos on YouTube, Tweet, and Facebook. This dataset provides information about the news content and user comments. 
However, these videos are collected from traditional video platform and social medias where editions of videos are not as prevalent and deep as those on emerging short video platforms, and thus could not reflect the latest challenge in fake news video detection. 
In addition, the publisher profiles are ignored in this dataset. 

\begin{table*}[]
\centering
\setlength{\abovecaptionskip}{.7em} 
\caption{Summary of datasets of fake news video detection. Metadata refers to basic statistics such as \# of likes/stars/comments.   
``-'' represents open-domain. Names of sources are abbreviated for simplicity (YT: YouTube, TW: Twitter, FB: Facebook, TT: TikTok, BB: Bilibili, DY: Douyin, KS: Kuaishou).  
}
\scalebox{0.8}
{
\begin{tabular}{lccccccc ccc}
\hline
\multirow{2}{*}{\hspace{3em}\textbf{Dataset}}        & \multicolumn{5}{c}{\textbf{Features}}                                                                       & \multirow{2}{*}{\textbf{Instances}}   & \multirow{2}{*}{\hspace{1em}\textbf{Domain}} & \multirow{2}{*}{\hspace{1em}\textbf{Language}} & \multirow{2}{*}{\hspace{0em}\textbf{Released}} & \multirow{2}{*}{\hspace{0.5em}\textbf{Source}
}\\ 
\cmidrule{2-6}
\textbf{}               & \textbf{Video}      & \textbf{Title}      & \textbf{Metadata} & \textbf{Comment} & \textbf{User}            & \textbf{
(fake/real)}            & \textbf{}       & \textbf{}     & \textbf{}                 \\
\hline
\cite{fvc}       & \checkmark                  & \checkmark                   & \checkmark                 & \checkmark                &                          & 2,916/2,090          & -               & 
En,Fr,Ru,Ge,Ar  &Y            & YT,TW,FB  \\
\cite{vavd}             & \checkmark                   & \checkmark                   & \checkmark                 & \checkmark                &                          & 123/423            & -               & En & Y            & YT                   \\
\cite{yt-pcancer}         & \checkmark                   &                     & \checkmark                 &                  &                          & 118/132            & prostate cancer & En & N             & YT                   \\
\cite{yt-covid}       & \checkmark                   & \checkmark                   &                   & \checkmark                &                          & 113/67            & COVID-19        & En & N             & YT                  \\
\cite{myvc}               & \checkmark                   & \checkmark                   &                   & \checkmark                &                          & 902/903            & -              & En & N             & YT                 \\
\cite{tt}           & \checkmark                   & \checkmark                   &                   &                  &                          & 226/665            & COVID-19       & En & N             & TT                   \\
\cite{bb}          & \checkmark                   &                     & \checkmark                 &                  &  & 210/490            & health         & Ch & N             & BB                \\
\textbf{FakeSV (ours)} & \textbf{\checkmark}          & \textbf{\checkmark}          & \textbf{\checkmark}        & \textbf{\checkmark}       & \textbf{\checkmark}              & \textbf{1,827/1,827} & \textbf{-}     & Ch & \textbf{Y}    & \textbf{DY, KS} \\
\hline
\end{tabular}}
\label{tab:datasets}
\vspace{-.9em}
\end{table*}

\subsubsection{Techniques.}
Far behind text-image fake news detection where many well-designed neural networks have been proposed to model different types of text-image correlations \cite{mm21}, most works oriented for video-based fake news detection are still in the preliminary stage of hand-crafted features. 
As the first work of detecting fake news videos, \cite{fvc} build an SVM classifier over features based on the video metadata, title linguistics, and comment credibility. 
\cite{yt-covid} use the tf-idf vectors of title and comments, and comments conspiracy to make a classification. \cite{yt-pcancer} firstly import emotion acoustic features and \cite{bb} convert audio into text to extract linguistic features.  Inspired by the rapid development of deep neural networks, \cite{vavd} use LSTM to model the comment features and concatenate them with hand-crafted features. 
\cite{myvc} propose a well-designed model that uses pre-trained models to extract features of video frames, title, and comments. The difference in topic distribution between title and comments is used to fuse these two modalities and a topic-adversarial classification is used to guide the model to learn topic-agnostic features for good generalization. 
\cite{tt} use the extracted speech text to guide the feature learning of visual frames, use MFCC \cite{mfcc} features to enhance the speech text, and then use a co-attention module to fuse the visual and speech information. 

Although existing works have made some achievements in fake news video detection, they only utilize partial modalities that is available in the experimental dataset for detection, which inevitably misses some important clues.
Moreover, as these methods are evaluated on different datasets, there is a lack of an effective comparison between them. 
Therefore, in this work, we propose a new multimodal detection model which considers all the involved modalities and reimplement these representative baseline methods on the proposed FakeSV dataset for a fair comparison.

\section{Dataset Construction}

\begin{figure}[htbp]
\setlength{\abovecaptionskip}{8pt}
\centering
	\includegraphics[width=.38\textwidth]{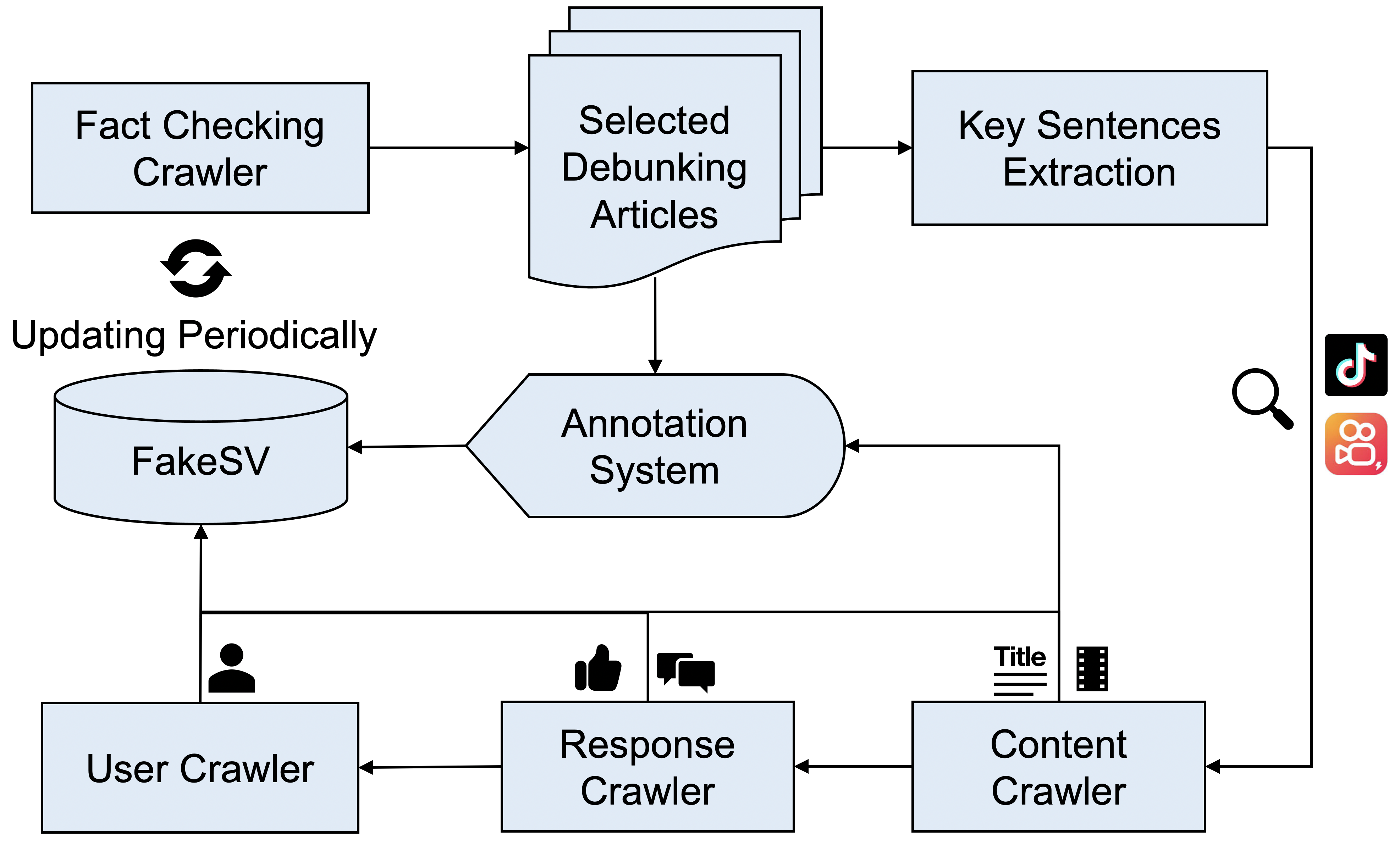}
	\caption{Flowchart of data construction process. }
	\label{fig:flowchart}
\end{figure}

Next, we describe the construction process of the FakeSV dataset as shown in Figure~\ref{fig:flowchart}.

\begin{table}[h]
\setlength{\abovecaptionskip}{.5em} 
\setlength{\belowcaptionskip}{-.3em}
\caption{Crawled fields in FakeSV.} 
\scalebox{0.9}{
\begin{tabular}{lp{.8\columnwidth}}
\hline
Category & \makecell[c]{Fileds}\\
\hline
Content & video, cover image, title, published time \\                                                                                          
Response     & \# of likes/stars/comments, top 100 comments (with reviewed time, \# of likes and \# of sub-comments) \\
Publisher    & info\_verified, info\_introduction, current IP location,  \# of fans/subscribes/likes/videos and top 100 published videos' covers    \\
\hline    
\end{tabular}}
\label{tab:fields}   
\vspace{-.8em}
\end{table}

\subsection{Data Collection}
To get as much fake news as possible and collect reliable ground truth labels for fake news, we utilized fact-checking websites to obtain the target events. We crawled 99,970 debunking articles from several official fact-checking sites between January 2011 and January 2022.
Articles without the word ``video" were dropped to increase the recall of retrieving video-form fake news.
To summarize news events from abundant fact-checking articles, we designed heuristic regular expressions to extract key sentences and 
removed duplicate news events using K-means clustering based on sentence representations of BERT \cite{bert}.
Then we paraphrased these key sentences in more general forms and finally obtained 854 event descriptions as search queries. 
Thereafter, we crawled relevant videos from two popular short video platforms in China, \ie Douyin\footnote{douyin.com} (the equivalent of TikTok in China) and Kuaishou\footnote{kuaishou.com}.
Besides video contents, we also crawled user responses and publisher profiles\footnote{Note that we only crawled the public information of the video's publisher for privacy reasons.} as shown in Table~\ref{tab:fields}.

\subsection{Data Annotation and Statistics}
\begin{figure}[tbp]
\setlength{\abovecaptionskip}{5pt}
\setlength{\belowcaptionskip}{-12pt}
	\centering
	\subfigure[Event]{
	\includegraphics[width=.15\textwidth]{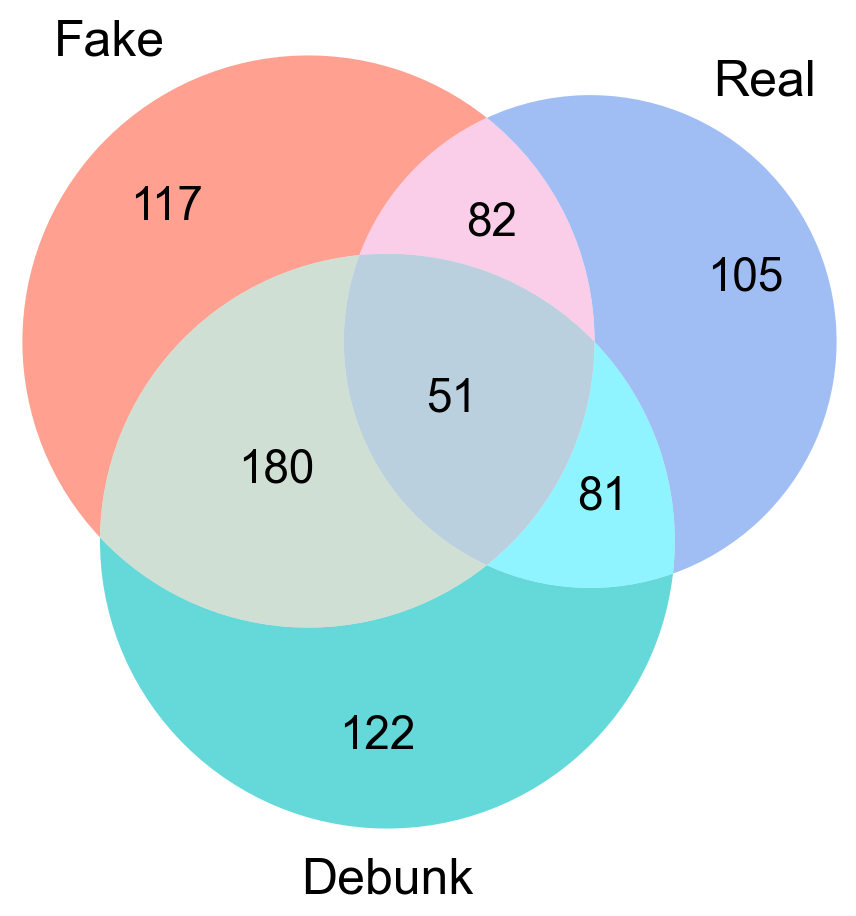}
	\label{fig:event}
	}
	\subfigure[Year]{\raisebox{0.15\height}{
	\includegraphics[width=.28\textwidth]{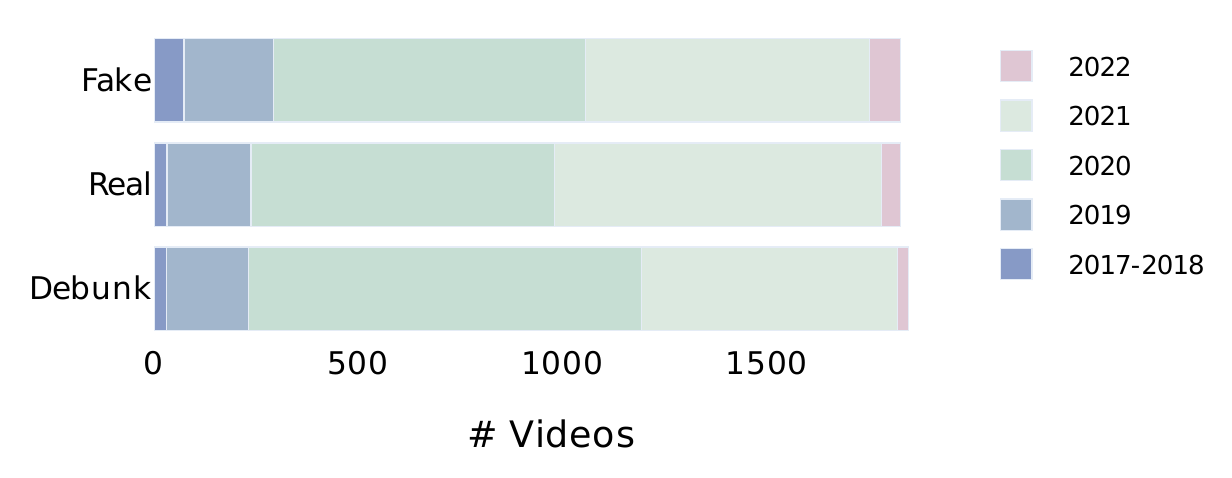}
	\label{fig:year}
	}}
	\caption{Distribution of FakeSV on event and year.}
	\label{fig:statistics}
\end{figure}

Although the searched queries are related to fake news that has been debunked, the retrieved videos are not always fake news videos. Therefore, we manually annotated\footnote{The details of annotation are introduced in the Appendix.} 
11,603 videos shorter than five minutes. The annotators were required to classify the given video into fake, real, debunked, and others (including useless and unverifiable videos).  
After balanced sampling, we ended up with 1,827 fake news videos, 1,827 real news videos, and 1,884 debunked videos under 738 events. 
Figure~\ref{fig:statistics} shows the event and time distribution  on the three classes of these videos.
Additionally, the percentage of real and fake news videos with comments is 75\% and 68\% respectively.

\section{Data Analysis}
To reveal the different behaviors between fake and real news videos, we provide some exploratory analyses from three perspectives, \ie news content, social context and propagation, to offer insights for detection.

\subsection{News Content}
\subsubsection{Text.} 
\begin{figure}[htbp]
\setlength{\abovecaptionskip}{0cm}
\setlength{\belowcaptionskip}{-12pt}
	\centering
	\subfigure[Fake]{
	\includegraphics[width=.22\textwidth]{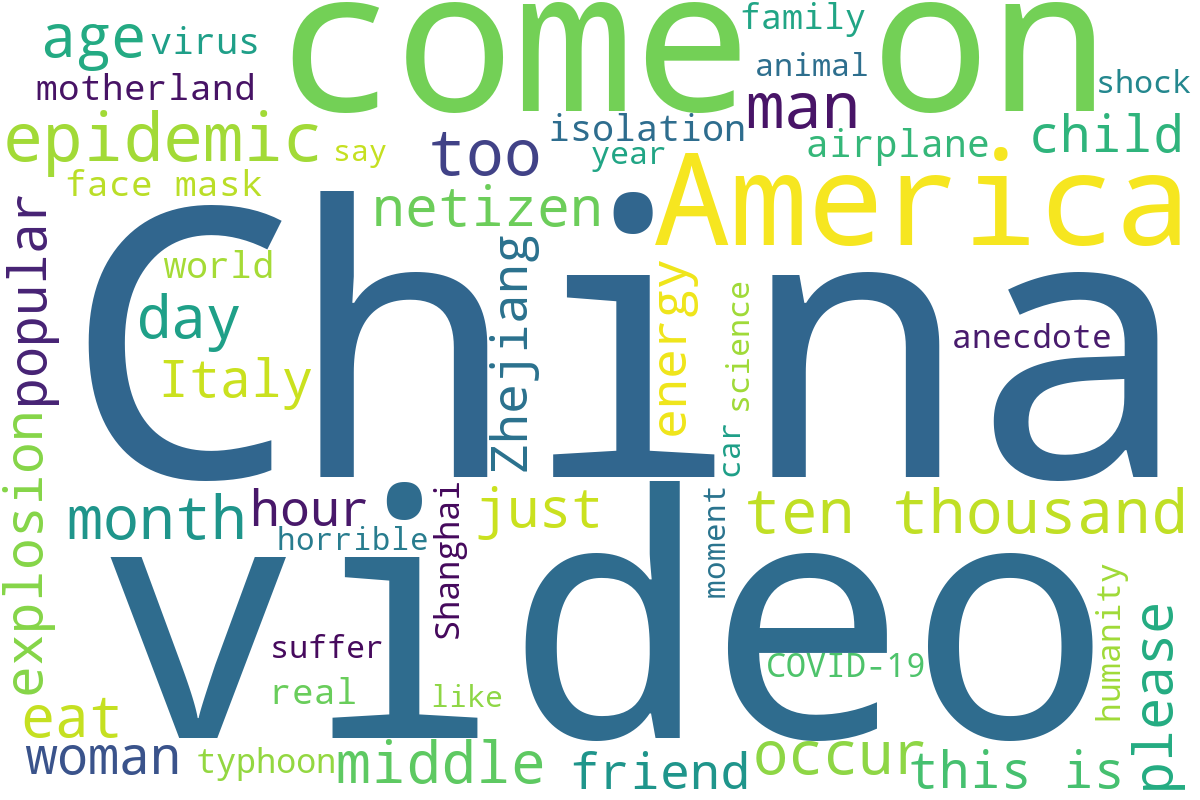}
	}
	\subfigure[Real]{
	\includegraphics[width=.22\textwidth]{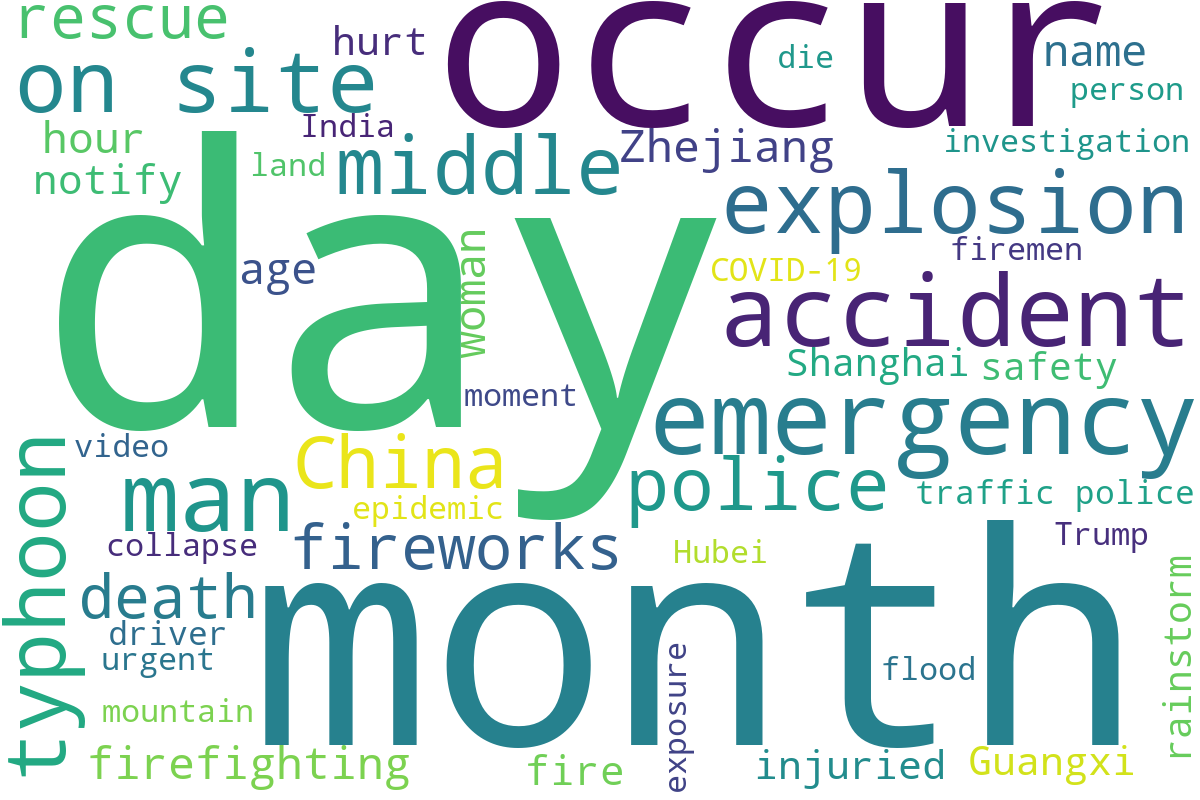}
	}
	\caption{Word cloud of video titles.}
	\label{fig:wordcloud}
\vspace{-.4em}
\end{figure}

Text is the dominant modality to detect fake news in the literature \cite{surveykdd2017}. We observe that fake news videos have shorter and more empty titles (see Appendix) providing less information compared with real news.
The word cloud of video titles in Figure~\ref{fig:wordcloud} shows that fake news titles emphasize the word ``video" much, prefer emotional and spoken words, and cover diverse topics. Differently, real news videos use more journalese and focus more on accidents and disasters.

\subsubsection{Video.} Image quality is considered to reflect originality and thus is used as an effective clue in text-image fake news detection \cite{icdm, chapter}. 
Following their work, we employ NIQE \cite{niqe} on video frames to indirectly measure video quality. Figure~\ref{fig:niqe} shows that fake news videos have lower quality than real news and contain videos with particularly poor quality.  

\subsubsection{Audio.} 
\begin{figure}[ht]
\vspace{-.4em}
\setlength{\abovecaptionskip}{3pt}
\setlength{\belowcaptionskip}{-12pt}
	\centering
	\begin{minipage}[t]{0.23\textwidth}
		\centering
		\includegraphics[width=\textwidth]{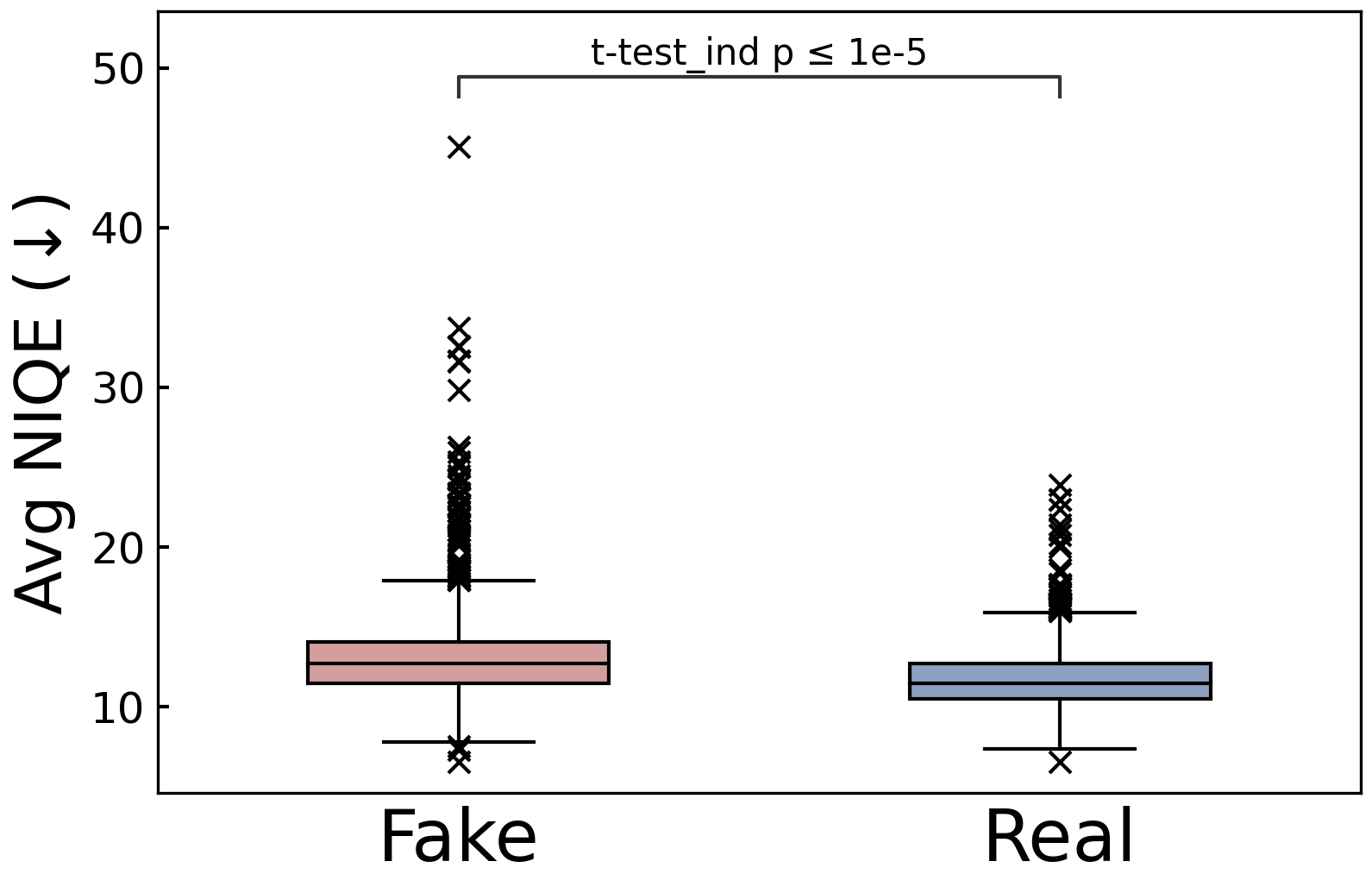}
		\caption{Frame quality. Lower values mean higher quality.}
		\label{fig:niqe}
	\end{minipage}
	\begin{minipage}[t]{0.23\textwidth}
		\centering
		\includegraphics[width=\textwidth]{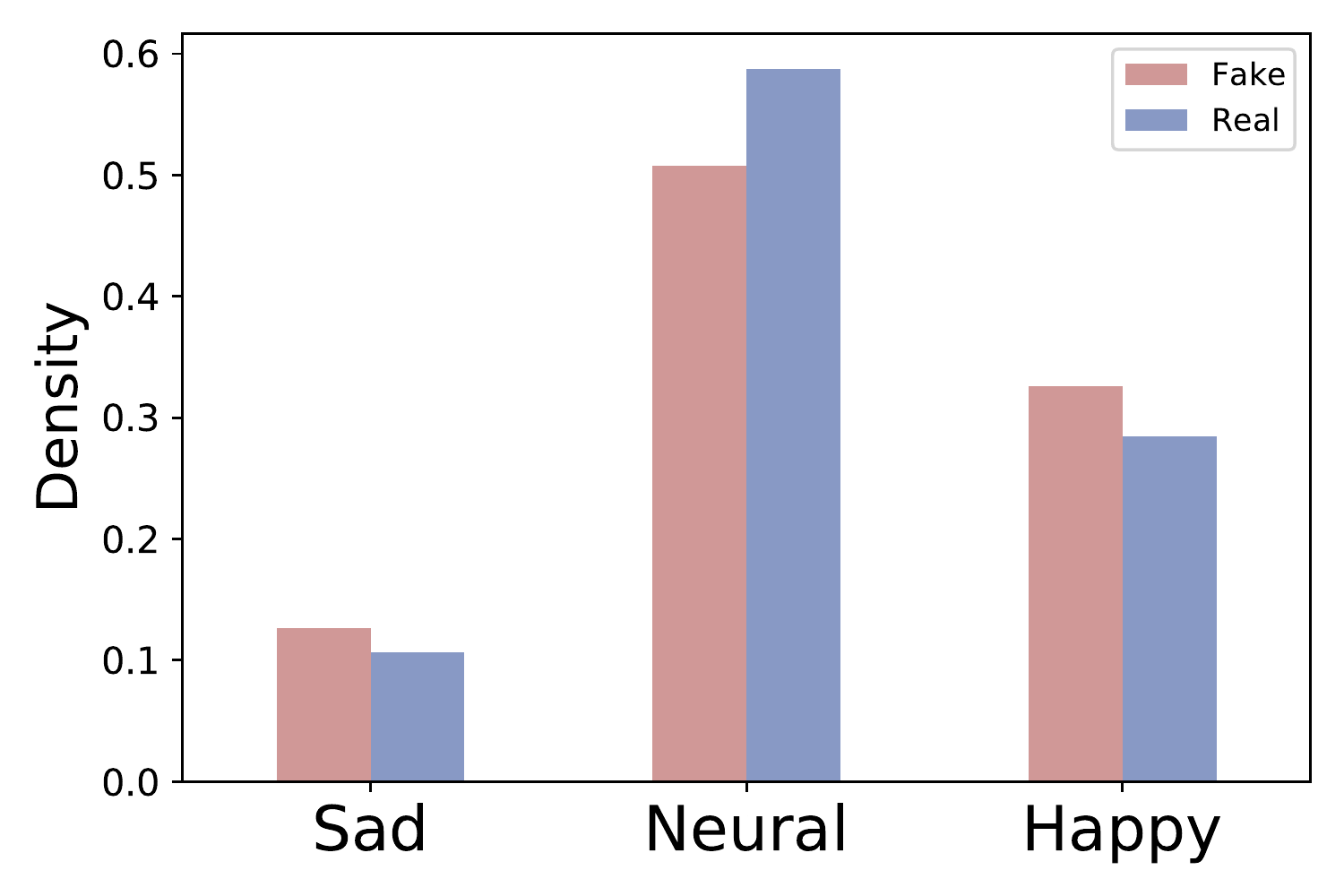}
		\caption{Audio emotion. }
		\label{fig:audioemo}
	\end{minipage}
\end{figure}

Emotion plays an important role in detecting fake news, and the emotional signals exist in the news text, news images or user responses \cite{dualemo}. Naturally, we analyze the speech emotion by the pre-trained wav2vec2 model \cite{wav2vec}. 
Figure~\ref{fig:audioemo} shows that the speech in fake news videos shows more obvious emotional preferences than real news.

\subsection{Social Context}
\subsubsection{Publisher Profiles. } 
\begin{figure}[b]
\vspace{-1em}
\setlength{\abovecaptionskip}{0cm}
\setlength{\belowcaptionskip}{-12pt}
	\centering
	\subfigure[Authority]{
	\includegraphics[width=.23\textwidth]{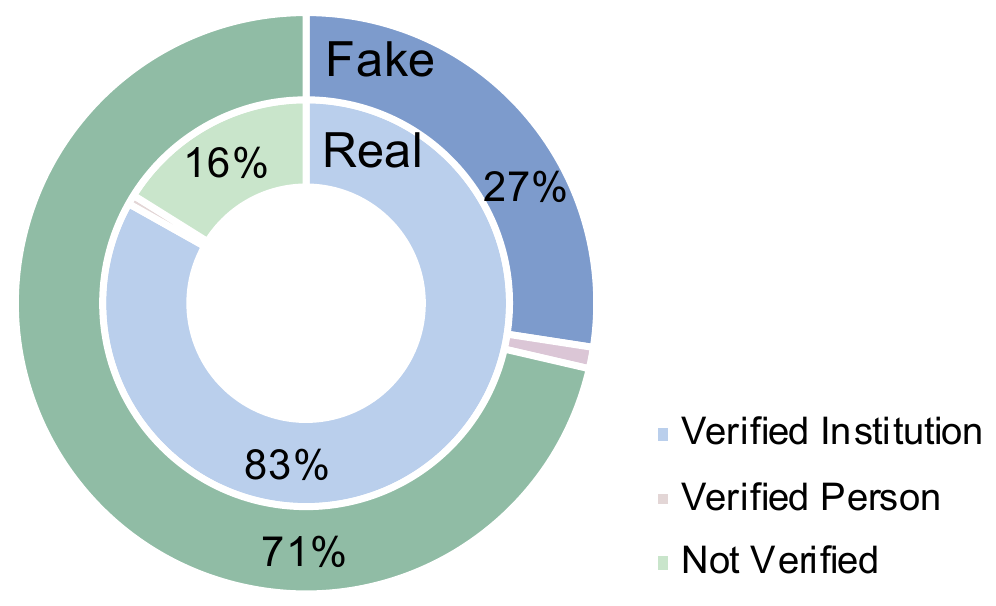}
	\label{fig:authority}
	}
	\subfigure[Statistics]{
	\includegraphics[width=.20\textwidth]{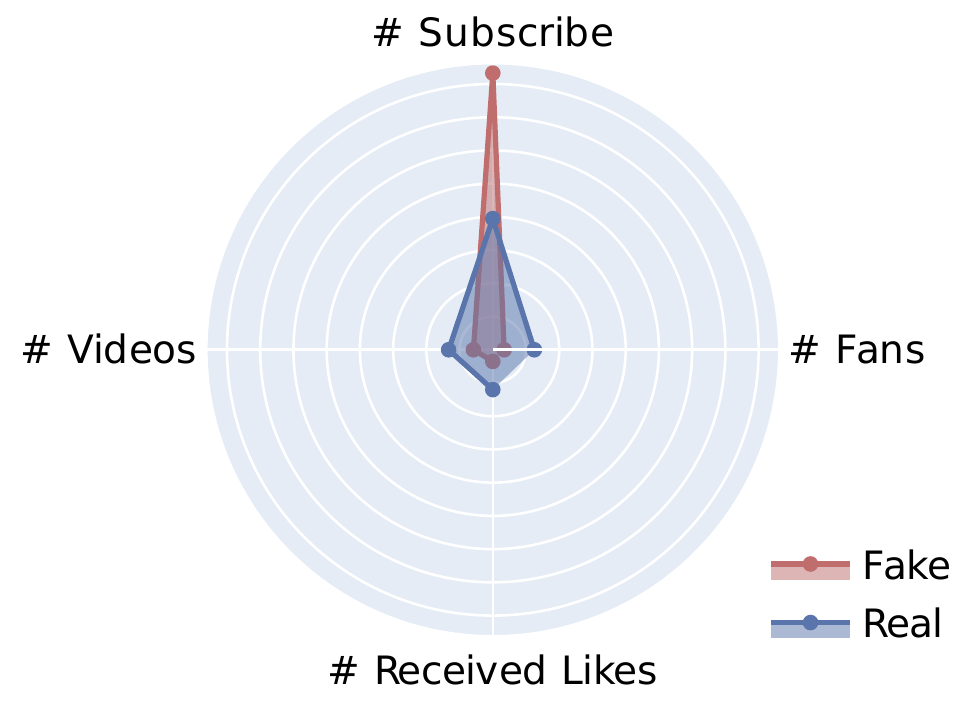} 
	\label{fig:radar}
	}
	\caption{Publisher's profile.}
	\label{fig:profile}
\end{figure}

User profiles on social media have been shown to be correlated with fake news \cite{profile}. As Figure~\ref{fig:authority} shows, most publishers of real news are verified accounts while most fake news publishers are not. Also, Figure~\ref{fig:radar} shows the average value of each user attribute after normalization. We could see that fake news publishers have more ``consuming" behaviors (subscribes) and less ``creating" behaviors (published videos, received likes, and fans) than real news publishers.

\subsubsection{User Responses.} 
\begin{figure}[t]
\setlength{\abovecaptionskip}{0cm}
\setlength{\belowcaptionskip}{-12pt}
	\centering
	\subfigure[Number of likes]{
	\includegraphics[width=.22\textwidth]{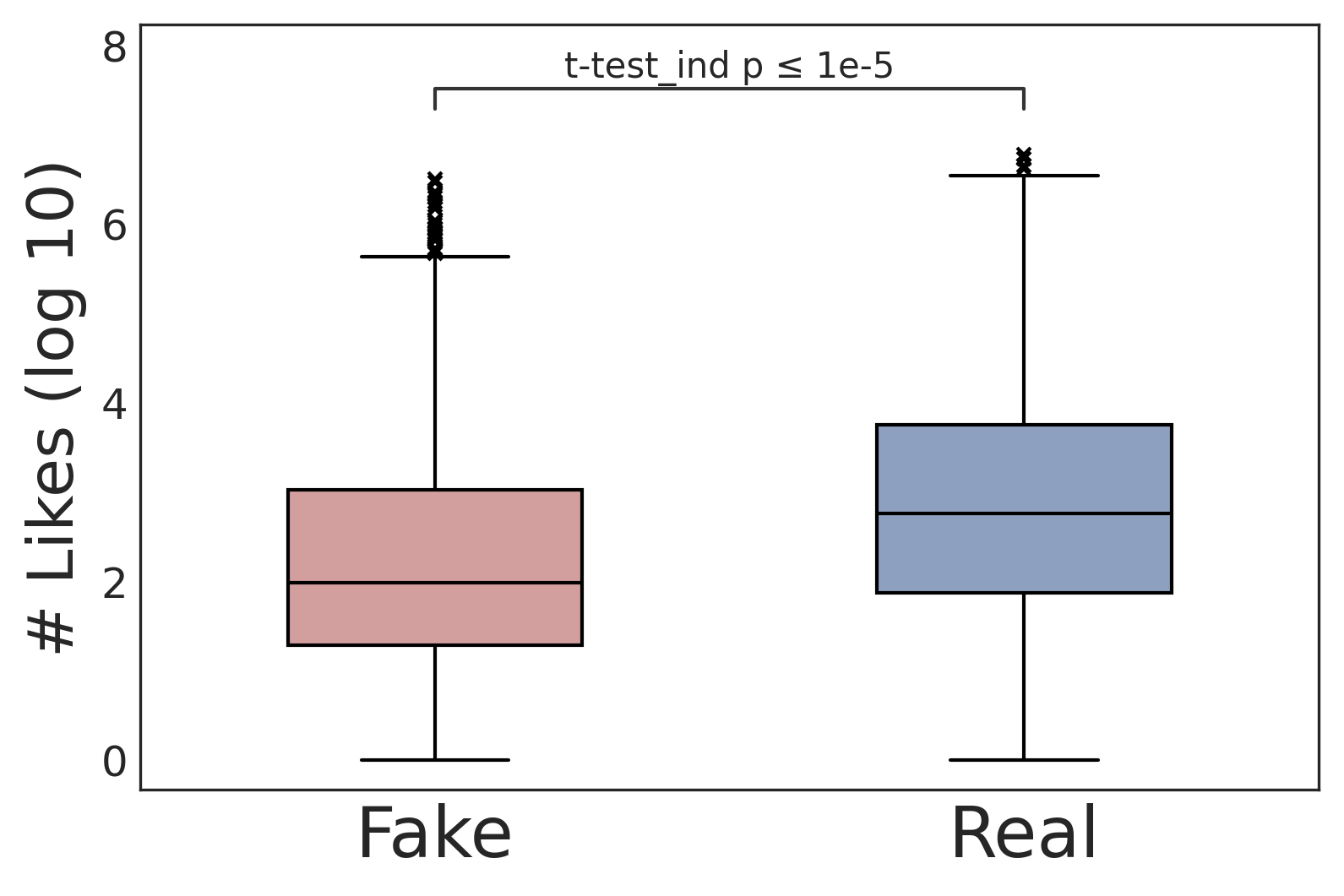}
	\label{fig:likes}
	}
	\subfigure[Relationship between the number of publisher fans and likes.]{
	\includegraphics[width=.22\textwidth,height=.15\textwidth]{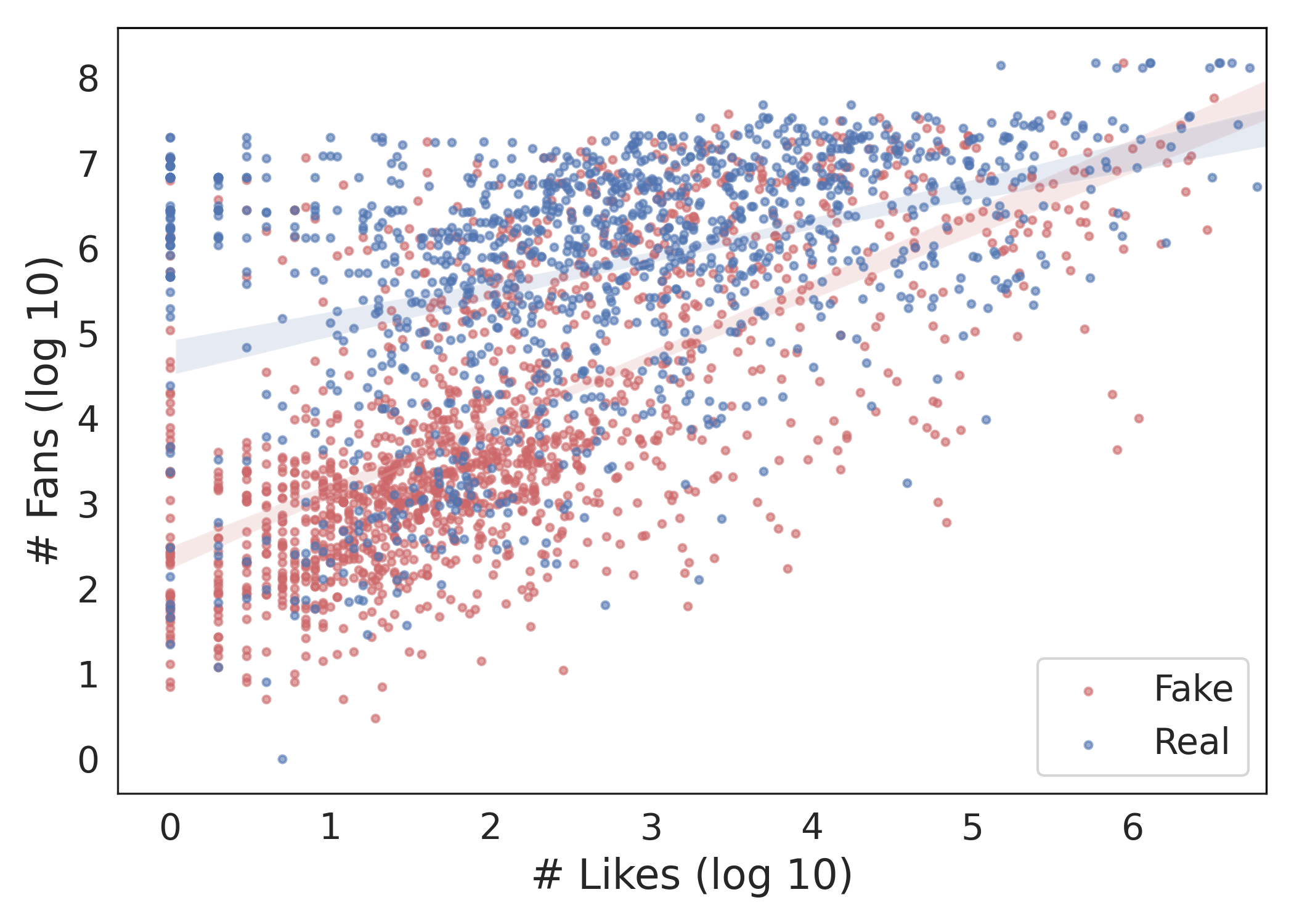}
	\label{fig:likes-fans}
	}
	\caption{User responses. }
\vspace{-.5em}
\end{figure}

We use the number of likes as an example to analyze user responses to fake and real news.
Figure~\ref{fig:likes} shows that real news videos receive more likes than fake news, which is intuitive considering that real news publishers have more fans.
To exclude the impact of account exposure, we analyze the relationship between the number of likes and the publisher's fans in Figure~\ref{fig:likes-fans}. We find that fake news videos receive more likes than real news when their publishers have a similar number of fans, which illustrates that fake news videos are more attractive than real news. 
In addition, the content of comments is also indicative in detecting fake news benefiting from crowd intelligence. 
According to our statistics, 18\% fake news videos receive doubtful comments (\eg ``Really?" and ``Fake!") while the ratio is only 4\% for real news.

\subsection{Propagation}
\subsubsection{Temporal Distribution.}
\begin{figure}[bp]
\vspace{-1.2 em}
\setlength{\abovecaptionskip}{0cm}
\setlength{\belowcaptionskip}{-12pt}
	\centering
	\subfigure[Temporal distributions in two event cases.]{
	\includegraphics[width=.23\textwidth]{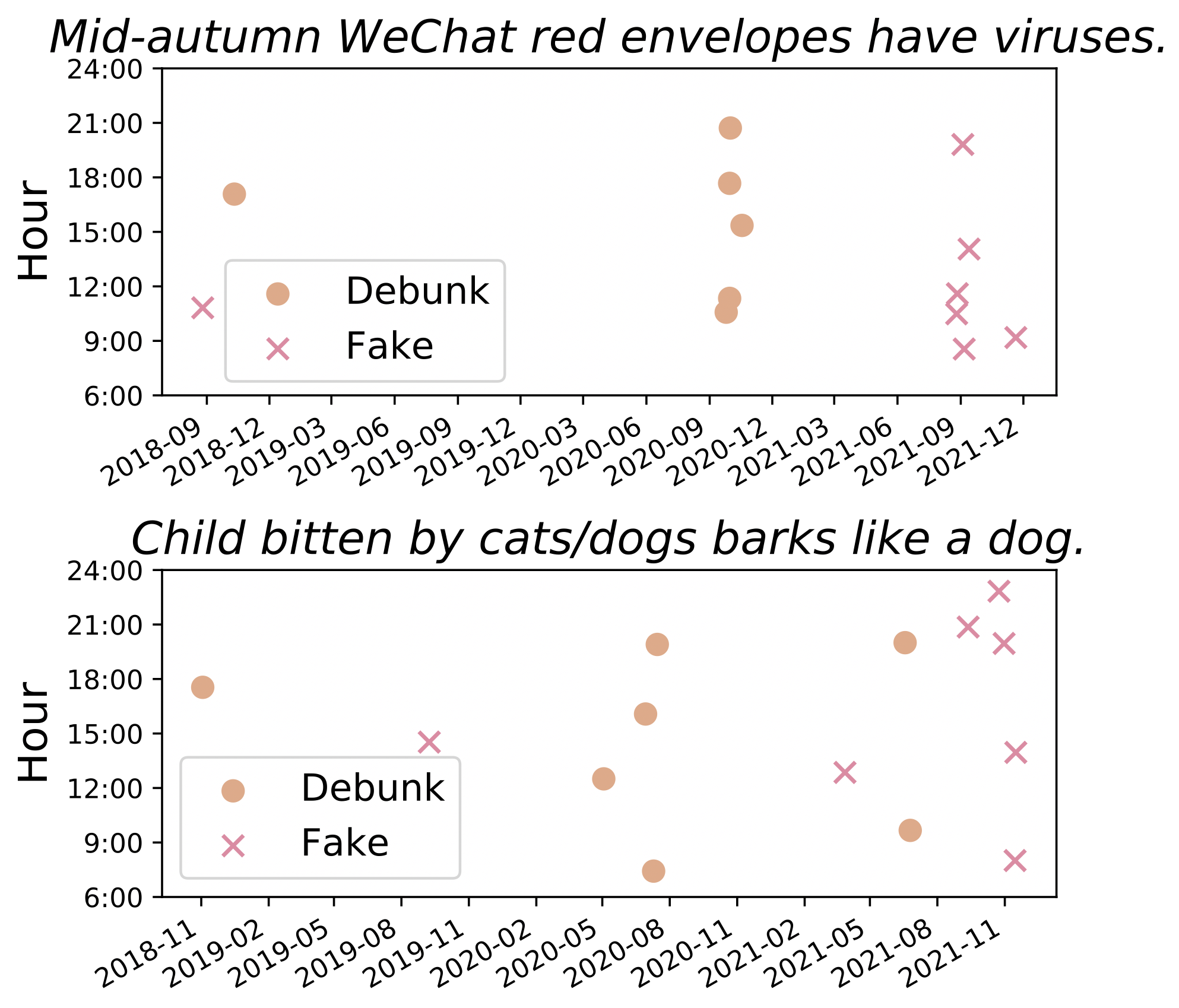}
	\label{fig:case-prop}
	}
	\subfigure[Video duplication. ]{\raisebox{0.15\height}{
	\includegraphics[width=.18\textwidth]{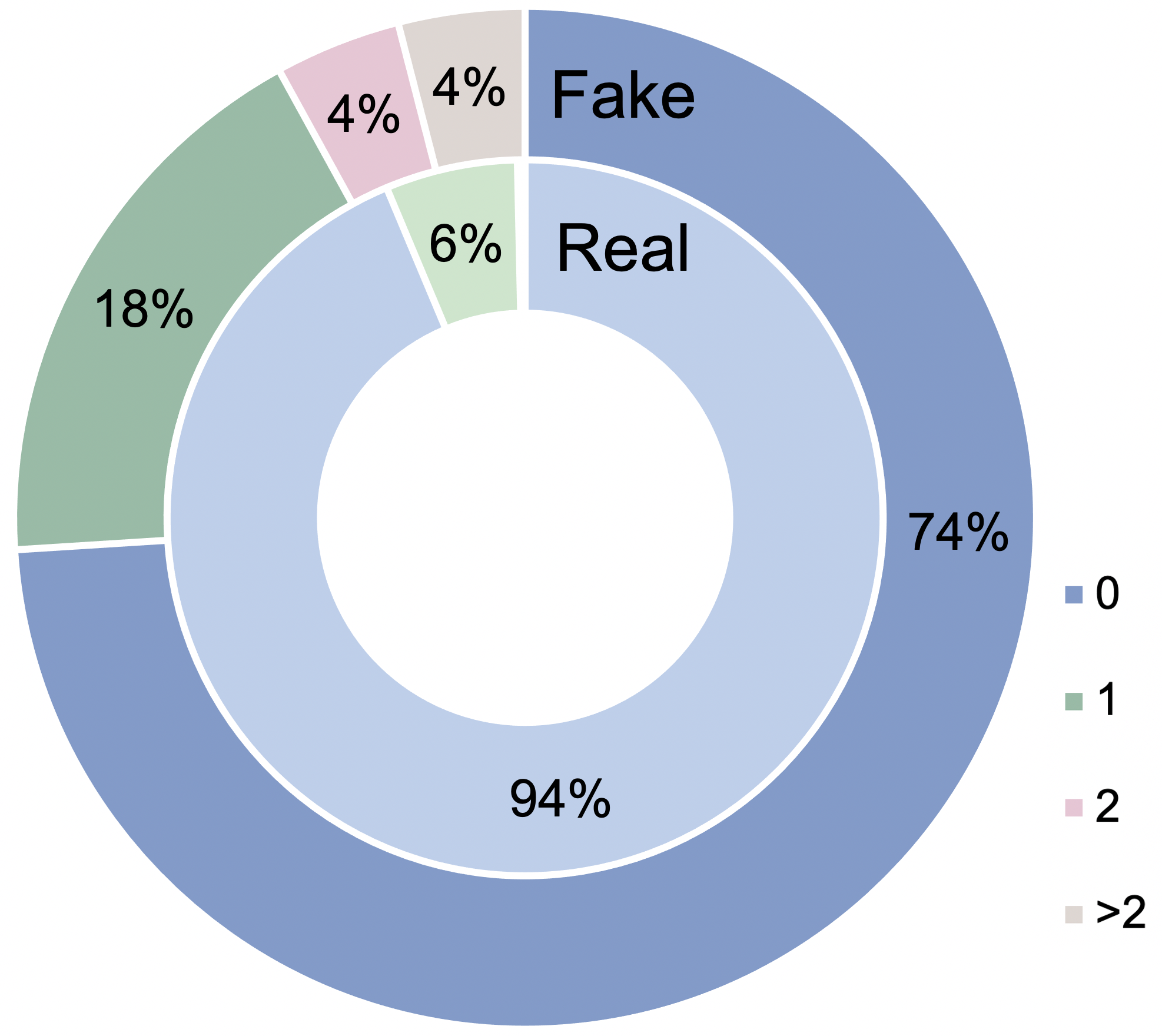} 
	\label{fig:videodup}
	}}
	\caption{Propagation analysis.}
	\label{fig:prof}
\end{figure}

To study the effect of debunking videos in preventing the propagation of fake news, we visualize their published time and find that fake news that has been previously debunked can still spread; this further highlights the importance of automatic detection of fake news videos. 
According to our statistics, for 434 events with debunking videos, 39\% of them have fake news videos emerged after the debunking videos were posted, especially the current or long-standing hot events. For example, as the first case in Figure~\ref{fig:case-prop} shows, the fake and debunking videos always occurred during the period of the mid-autumn festival (around September); they occurred concomitantly in long-standing hot events (the second case in Figure ~\ref{fig:case-prop}).

\subsubsection{Video Duplication.} Different from traditional social medias, there is no explicit propagation behaviors like retweeting on short video platforms. 
With the convenience of video editing functions provided by these platforms, people tend to edit and re-upload the videos, usually with no mention of the source. 
Therefore, we employ the pHash cluster algorithm \cite{phash} on the cover images to indirectly analyze the video duplication. 
As Figure~\ref{fig:videodup} shows, fake news videos have higher repetition while real news videos are more diverse. 
This phenomenon is due to the fact that real events will receive various images/videos taken by different witnesses \cite{tmm}.

\section{Method}
\subsection{Model Overview}
\begin{figure}[htbp]
\setlength{\abovecaptionskip}{10pt}
\setlength{\belowcaptionskip}{-12pt}
	\centering
	\includegraphics[width=.5\textwidth]{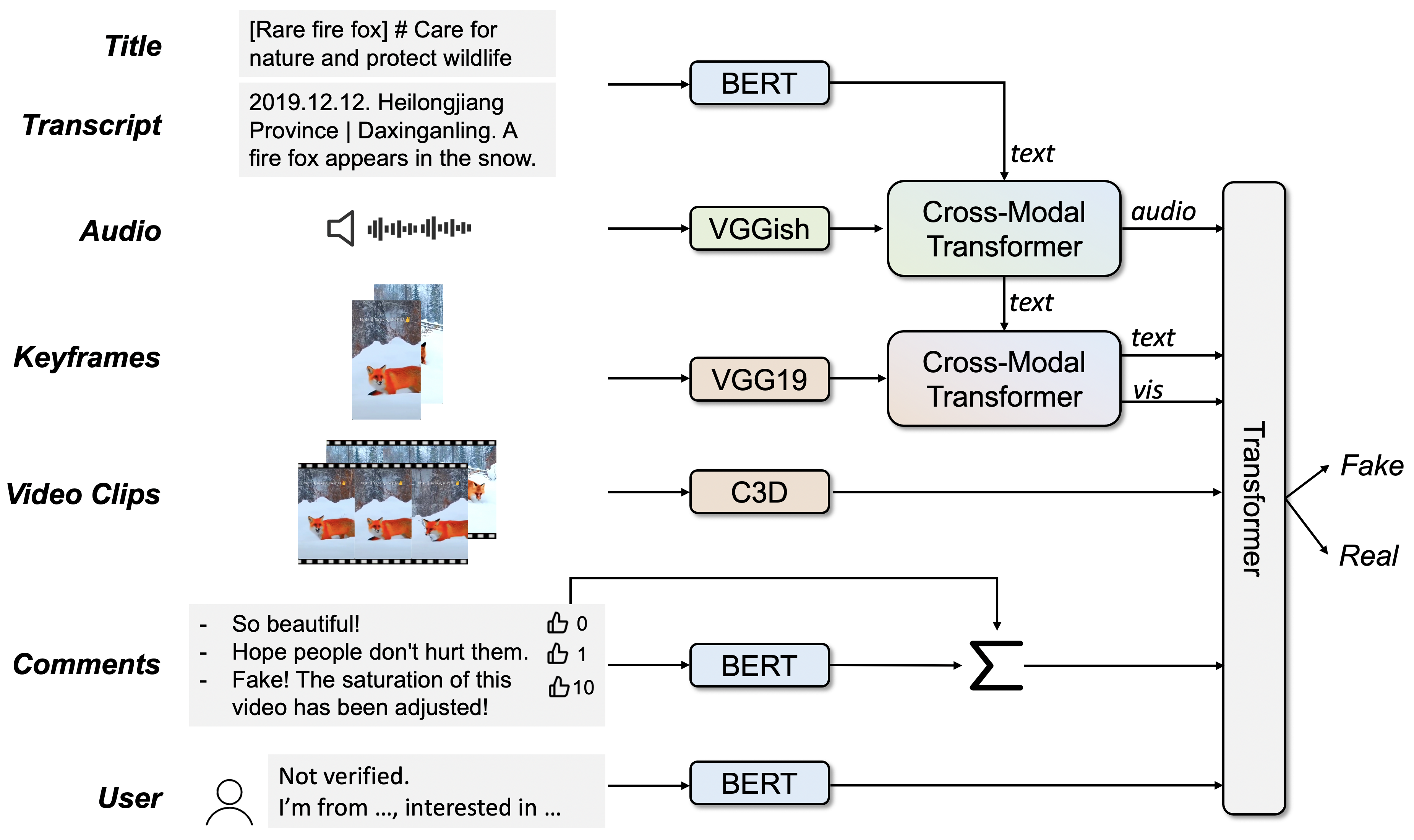}
	\caption{Architecture of the proposed framework SV-FEND. }
	\label{fig:framework}
\end{figure}

Figure~\ref{fig:framework} illustrates the overall architecture of the proposed SV-FEND. In order to model the diverse characteristics of the input news video, we first extract features of multiple modalities, including the text, audio, keyframes, video clips, comments, and user.  Next, we employ two cross-modal transformer layers to model the correlations between the text and the audio and keyframes. Another transformer layer is further used to dynamically fuse the multimodal features of news content and social context for the final classification.   

\subsection{Multimodal Feature Extraction}
\subsubsection{Title\&Transcript. }
For video-form news, the embedded text in the video is usually more complete than the title of which the length is limited. Therefore, we extract the video transcript and concatenate it with the title. The composed text is further fed into the pre-trained BERT to extract the textual features 
$\bm{H}_T=[\bm{w}_1,..., \bm{w}_l]$, where $\bm{w}_i$ represents the feature of the $i$-th word and $l$ is the text length. 

\subsubsection{Audio. } 
As a new imported modality compared to the text-image fake news, the audio modality includes speech, environment sound, and background music. Thus it provides not only semantics but also distinctive information in understanding the emotion of speakers and even the intent of the publisher. Specifically, we split the audio from the original video, and then use the pre-trained VGGish model \cite{vggish} to extract the audio features 
$\bm{H}_A=[\bm{a}_1,..., \bm{a}_n]$, where $n$ is the number of audio frames. 

\subsubsection{Video. } 
We represent a video both at the frame and clip levels to obtain spatiotemporal and multi-granularity information for detection \cite{wu2015modeling}. 
Taking Fig. 1(a) as an example, tampered and untampered regions have inconsistent compression rates (frame level) and motion trajectories (clip level).
At the frame level, we use the pre-trained VGG19 \cite{vgg19} model to extract the static visual feature $\bm{H}_I=[\bm{i}_1,..., \bm{i}_m]$, where $m$ is the number of frames. 
We also take 16 subsequent frames centered at each time step as a video clip, and use the pre-trained C3D model \cite{c3d} to extract the motion features $\bm{H}_V=[\bm{v}_1,..., \bm{v}_m]$. An average operation is applied to obtain the aggregated motion feature $\bm{x}_V$. 

\subsubsection{Comments. }
Similar to text modality, we use the pre-trained BERT to extract the features of each comment, that is $\bm{H}_C=[\bm{c}_1,..., \bm{c}_k]$, where $k$ is the number of comments.
Moreover, we use the number of likes to measure the importance of each comment.
Specifically, the feature of comments is computed as:
\begin{equation}
	\bm{x}_C = \sum_{j=1}^{k} \frac{l_j+1}{\sum_{j=1}^k l_j+k} \bm{c}_j,
\end{equation}
where $l_j$ is the number of likes towards the $j$-th comments. 

\subsubsection{User. } We concatenate the self-written and verified introduction of the publisher and feed it into the pre-trained BERT. The embedding of the $[CLS]$ token is extracted as the user features $\bm{x}_U$.

\subsection{Multimodal Feature Fusion}
The correlations between different modalities could help understand video semantics. For example, the acoustic characteristics of the audio modality often contain useful hints (such as adjusting the volume and tone) in capturing the key information in the speech text. The text could help select important ones from complex visual frames, and vice versa. Therefore, we use two cross-modal transformers to model the mutual enhancement between text and other modalities. 

The first cross-modal transformer in Figure~\ref{fig:framework} takes the textual feature $\bm{H}_T$ and the audio feature $\bm{H}_A$ as input, and the audio-enhanced textual feature is further fed into the second cross-modal transformer to interact with the frame feature $\bm{H}_I$.
Specifically, we use a two-stream co-attention transformer like \cite{vilbert} to process the multimodal information simultaneously, where the queries from each modality are passed to the other modality’s multi-headed attention block.
Finally, we obtain the textual feature enhanced by audio and visual frames $\bm{H}_{T\gets A,I}$, text-enhanced audio feature $\bm{H}_{A\gets T}$, and text-enhanced frame feature $\bm{H}_{I\gets T}$. We average them and then obtain the features $\bm{x}_T, \bm{x}_A, $ and $\bm{x}_I$. 

Till now, we have extracted six features to represent the input video from different perspectives, and used two cross-modal transformers to model the correlations between different modalities in news content. 
Thereafter, we utilize self-attention \cite{trm} to model the correlations between features from news content and social context. 
Specifically, we concatenate the obtained six features into a feature sequence and feed it into a standard transformer layer, finally obtaining the multimodal fused feature $\bm{x}_m$.

\begin{table*}[]
\centering
\caption{Performance (\%) comparison between the proposed model and the baseline models. We report the mean and standard deviation of the five-fold cross-validation. Data formats are abbreviated for simplicity (M: Metadata, T: Title, Tr: Transcript, C: Comment, F: Keyframe, V: Video clip, A: Audio).}
\centering
\begin{tabular}{ccccccc}
\hline
\textbf{Modality} & \textbf{Data} & \textbf{Method} & \textbf{Acc.} & \textbf{F1} & \textbf{Prec.} & \textbf{Recall} \\
\hline
 & M & hc feas+SVM & 67.98$_{\pm1.12}$ & 67.92$_{\pm1.15}$ & 68.11$_{\pm1.08}$ & 67.98$_{\pm1.12}$ \\
 \cline{2-7}
& &  hc feas+SVM & 64.27$_{\pm2.08}$ & 64.15$_{\pm2.14}$ & 64.63$_{\pm2.27}$ & 64.41$_{\pm2.23}$ \\
\multirow{-3}{*}{Social}& \multirow{-2}{*}{C}&  BERT+Att & 62.87$_{\pm3.52}$ & 60.62$_{\pm6.91}$ & 63.33$_{\pm4.57}$ & 62.13$_{\pm4.99}$ \\
\hline
& & hc feas+SVM & 70.22$_{\pm2.10}$ & 70.31$_{\pm2.21}$ & 70.37$_{\pm2.18}$ & 70.33$_{\pm2.20}$ \\
& & Text-CNN & 74.66$_{\pm2.31}$ & 74.62$_{\pm2.30}$ & 74.79$_{\pm2.41}$ & 74.66$_{\pm2.31}$ \\

\multirow{-3}{*}{Text} & \multirow{-3}{*}{T\&Tr} &  BERT & 76.82$_{\pm3.33}$ & 76.80$_{\pm3.34}$ & 76.89$_{\pm3.33}$ & 76.82$_{\pm3.33}$ \\
\hline
& &  VGG19+Att & 69.40$_{\pm2.64}$ & 69.33$_{\pm2.57}$ & 69.64$_{\pm2.89}$ & 69.40$_{\pm2.64}$ \\
& \multirow{-2}{*}{F} &  Faster R-CNN+Att & 70.69$_{\pm2.80}$ & 70.58$_{\pm2.76}$ & 71.03$_{\pm2.99}$ & 70.69$_{\pm2.80}$ \\
\cline{2-7}
\multirow{-3}{*}{Visual}& V &  C3D+Att & 69.05$_{\pm1.71}$ & 68.93$_{\pm1.66}$ & 69.36$_{\pm1.87}$ & 69.05$_{\pm1.71}$ \\
\hline
& &  emo feas+SVM & 60.64$_{\pm1.22}$ & 60.61$_{\pm1.24}$ & 60.67$_{\pm1.20}$ & 60.64$_{\pm1.22}$ \\
\multirow{-2}{*}{Audio}& \multirow{-2}{*}{A} & VGGish & 66.78$_{\pm1.12}$ & 66.63$_{\pm1.13}$ & 67.07$_{\pm1.20}$ & 66.78$_{\pm1.12}$ \\

\hline
\multirow{5}{*}{Multimodal}& M, Tr, A  & \cite{yt-pcancer} & 68.64$_{\pm2.01}$ & 68.01$_{\pm1.99}$ & 70.24$_{\pm2.42}$ & 68.64$_{\pm2.01}$ \\
& T, C&  \cite{yt-covid} & 71.45$_{\pm2.43}$ & 71.45$_{\pm2.44}$ & 71.47$_{\pm2.42}$ & 71.45$_{\pm2.43}$ \\
& T, C, F&  \cite{myvc} & 75.07$_{\pm0.38}$ & 75.04$_{\pm0.38}$ & 75.18$_{\pm0.38}$ & 75.07$_{\pm0.38}$ \\
& Tr, F, A &  \cite{tt} & 75.04$_{\pm3.28}$ & 75.02$_{\pm3.29}$ & 75.11$_{\pm3.27}$ & 75.04$_{\pm3.28}$ \\
& \cellcolor{gray!25}ALL &  \cellcolor{gray!25}SV-FEND(ours) & \cellcolor{gray!25}\textbf{79.31$_{\pm2.75}$} & \cellcolor{gray!25}\textbf{79.24$_{\pm2.79}$} & \cellcolor{gray!25}\textbf{79.62$_{\pm2.60}$} & \cellcolor{gray!25}\textbf{79.31$_{\pm2.75}$} \\

\hline
\end{tabular}
\label{tab:comparison}
\end{table*}

\subsection{Classification}
We use a fully connected layer with softmax activation to project the multimodal feature vector $\bm{x}_m$ into the target space of two classes: real and fake news videos, and gain the probability distributions:
\begin{equation}
	\bm{p} = {\rm softmax}(\bm{W}_b \bm{x}_m + \bm{b}_b),
\end{equation}
where $\bm{p} = [p_0, p_1]$ is the predicted probability vector with $p_0$ and $p_1$ indicate the predicted probability of label being 0 (real news video) and 1 (fake news video) respectively. $\bm{W}_b$ is the weight matrix and $\bm{b}$ is the bias term. 
Thus, for each news post, the goal is to minimize the binary cross-entropy loss function as follows,
\begin{equation}
	\mathcal{L}_p=-[(1-y)\log p_0 + y \log p_1],
\end{equation}
where $y \in \{0,1\}$ denotes the ground-truth label. 

\section{Experiments}
 
\subsection{Baseline Methods}
To establish a comprehensive benchmark, we experiment with multiple representative methods that utilize single or multiple modalities on FakeSV. 

\textbf{Social Modality}: 
For Metadata (\textbf{M}), we combine the \textbf{h}and-\textbf{c}rafted features (\textbf{hc feas}) of metadata proposed in \cite{yt-pcancer, bb} and classify them by  \textbf{S}upport \textbf{V}ector \textbf{M}achine (\textbf{hc feas+SVM}). 
For Comment (\textbf{C}), we use the \textbf{hc feas} about comments proposed in \cite{vavd, yt-covid, bb} and \textbf{SVM} as classifier. Also, we use \textbf{BERT} to extract the features of multiple comments and use the \textbf{Attention} mechanism to fuse them for classification. 

\textbf{Text Modality}-Title (\textbf{T})\&Transcript (\textbf{Tr}):
We use \textbf{hc feas} about the title and transcript proposed in \cite{fvc, vavd, yt-covid} and \textbf{SVM} as classifier. Also, we adopt \textbf{Text-CNN} \cite{textcnn} and \textbf{BERT} as textual encoders with an MLP for classification. 

\textbf{Visual Modality}: 
For keyframe (\textbf{F}), we use two visual encoders used in existing multimodal baselines, that is \textbf{VGG19} and \textbf{Faster R-CNN} \cite{fasterrcnn}, to extract the visual features of multiple frames.
For video clip (\textbf{V}), we use the pre-trained \textbf{C3D} to extract the motion features. 
The \textbf{Attention} mechanism is used to fuse features of different time steps for classification.  

\textbf{Audio Modality (A)}:  We extract the emotion audio features (\textbf{emo feas}) proposed in \cite{yt-pcancer} and use \textbf{SVM} to classify. Also, we use the pre-trained \textbf{VGGish} model to extract the acoustic features for classification.

\subsubsection{MultiModality: } We use four existing state-of-the-art methods as multimodal baselines: 
1) \textbf{\cite{yt-pcancer}} use the linguistic features from the speech text, acoustic emotion features, and user engagement features and a linear kernel SVM to distinguish the real and fake news videos. 
2) \textbf{\cite{yt-covid}} extract tf-idf vectors from the title and the first hundred comments and use traditional machine learning classifiers including logistic regression and SVM. 
3) \textbf{\cite{myvc}} use the topic distribution difference between title and comments to fuse them, and concat them with the visual features of keyframes. An adversarial neural network is used as an auxiliary task to extract topic-agnostic multimodal features for classification. 
4) \textbf{\cite{tt}} use the extracted speech text to guide the learning of visual object features, use MFCC features to enhance the speech textual features, and then use a co-attention module to fuse the visual and speech information for classification.

\subsection{Evaluation}
Prior experiments have shown that the detection performance varies a lot under different data splits.
To ensure fairness of experiments, our evaluations are conducted by doing five-fold cross-validation with accuracy (Acc.), macro precision (Prec.), macro recall (Recall), and macro F1-score (F1) as evaluation metrics.
For each fold, the dataset is split as training and testing sets at the event level with a ratio of 4:1, ensuring that there is no event overlap among different sets. 

\subsection{Experimental Results}

\subsubsection{Performance Comparison.} Table~\ref{tab:comparison} shows the results of the baseline models and the proposed model, from which we can draw the following observations: 
1) SV-FEND performs much better than the other methods, which validates that SV-FEND can effectively capture important multimodal clues to detect fake news videos. 
2) According to the best-performing methods, the discriminability of different data could be sorted as: title\&transcript \textgreater keyframe \textgreater video clip \textgreater metadata \textgreater audio \textgreater comments. 
3) The best performance in terms of average accuracy on FakeSV is less than 0.8, which is lower than the accuracy achieved on other popular multimodal fake news datasets \citep{mm17,mediaeval16} in which the best accuracy is higher than 0.9. This further demonstrates the challenges of this dataset. 

\begin{table}[t]
\setlength{\abovecaptionskip}{.7em} 
\setlength{\belowcaptionskip}{-.7em}
\centering
\caption{Ablation study on different modalities. The standard deviation values are ignored for simplicity. }
\begin{tabular}{lcccc}
\hline
\multicolumn{1}{c}{\textbf{Method}} & \textbf{Acc.}   & \textbf{F1}     & \textbf{Prec.}  & \textbf{Recall} \\
\hline
\textbf{SV-FEND}     & \textbf{79.31}  & \textbf{79.24}  & \textbf{79.62}  & \textbf{79.31}  \\
\hline
w/o News Content & 74.89 & 74.57 & 76.34 & 74.89 \\
\quad w/o Text     & 75.37         & 75.11         & 76.54         & 75.37        \\
\quad\quad w/o Title          & 77.28         & 77.23         & 77.53          & 77.28        \\
\quad\quad w/o Transcript    & 77.89          & 77.86         & 78.01    & 77.89         \\
\quad w/o Visual 			& 77.97          & 77.92         & 78.20          & 77.97       \\
\quad\quad w/o Keyframes   & 78.49          & 78.43          & 78.77         & 78.49        \\
\quad\quad w/o Video Clips & 78.95          & 78.93         & 79.07         & 78.95         \\
\quad w/o Audio         & 78.95        & 78.93        & 79.03          & 78.95          \\
\hline
w/o Social Context & 78.62 & 78.54 & 78.98 & 78.62\\
\quad w/o User             & 78.76          & 78.66         & 79.26          & 78.76         \\
\quad w/o Comment       & 79.09         & 79.02        &  79.45          & 79.09          \\
\hline
\end{tabular}
\label{tab:ablation}
\vspace{-.3em}
\end{table}

\subsubsection{Analysis on Different Modalities.}
We conduct ablation experiments to analyze the importance of each modality in detecting fake news videos. From Table~\ref{tab:ablation}, we can see that 1) All modalities in SV-FEND are useful for achieving its best performance. 2) News content is more effective than social context in detecting fake news. 
3) Different data under the same modality have homogeneity and complementarity, such as title and transcript, and keyframes and video clips. 
4) Comments play the least role in the detection, which maybe results from data sparsity.

\begin{table}[]
\centering
\setlength{\abovecaptionskip}{.6em} 
\caption{Performance comparison under temporal split. }
\begin{tabular}{ccccc}
\hline
\textbf{Method} & \textbf{Acc.}                    & \textbf{F1}                      & \textbf{Prec.}                   & \textbf{Recall}                  \\
\hline
\cite{yt-pcancer}  &  71.89 & 71.29 & 73.88 & 71.89 \\                             
\cite{yt-covid}    &      75.58  &      75.50  & 75.92   &   75.58   \\
\cite{myvc}   &   78.32   &  78.31   &       78.37   &      78.32 \\
\cite{tt}   &   74.45  &  74.39  &  74.67   &  74.45 \\
SV-FEND(ours)     & \textbf{81.05} & \textbf{81.02} & \textbf{81.24} & \textbf{81.05} \\
\hline
\end{tabular}
\label{tab:temporal}
\end{table}

\subsubsection{Performance on Temporal Split.}
In real-world scenarios, when a check-worthy news video emerges, we only have the previously-emerging data to train the detector.
Therefore, we provide a temporal split in chronological order with a ratio of 70\%:15\%:15\%, to evaluate the ability of models to detect future fake news videos. Table~\ref{tab:temporal} validates the superiority of the proposed SV-FEND model in the temporal split. Also, we observe that the performance in the temporal split is higher than that in the event split, which is due to the existence of long-standing fake news events.

\subsection{Case Studies}
\begin{figure}[tbp]
\setlength{\abovecaptionskip}{0cm}
\setlength{\belowcaptionskip}{-10pt}
	\centering
	\subfigure[Detected]{
	\includegraphics[height=.22\textwidth]{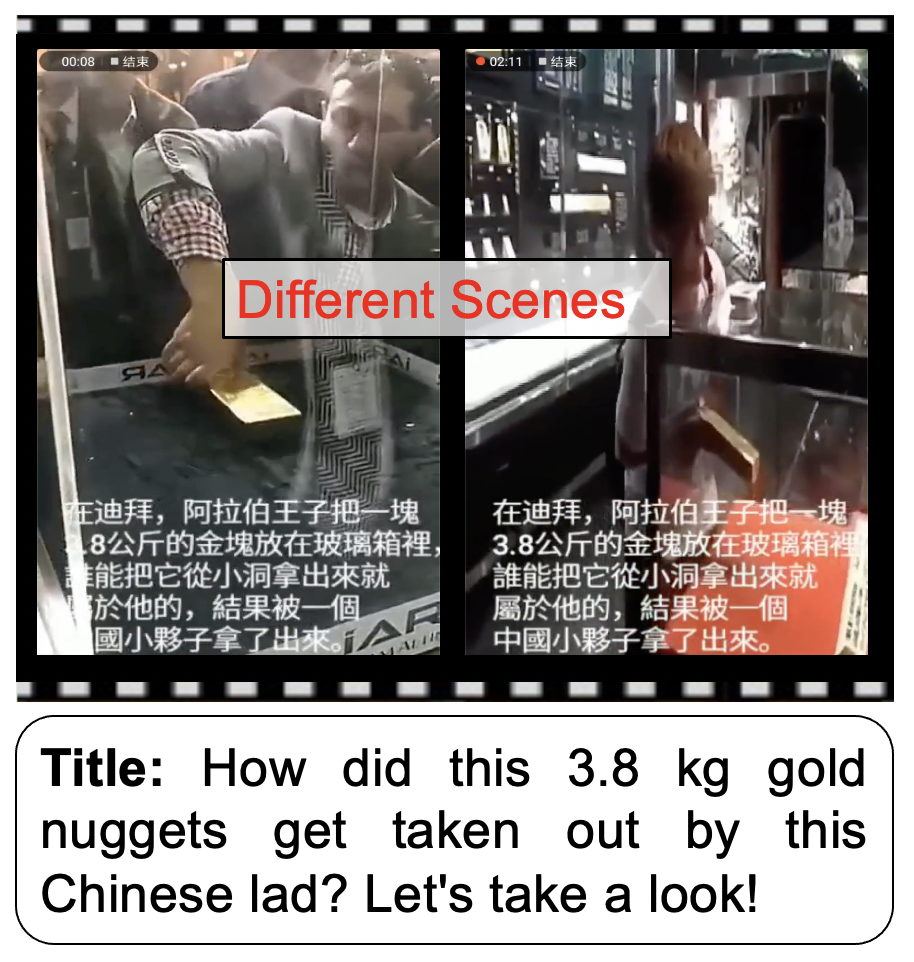}
	\label{fig:case-succeed}
	}\
	\subfigure[Missed]{
	\includegraphics[height=.22\textwidth]{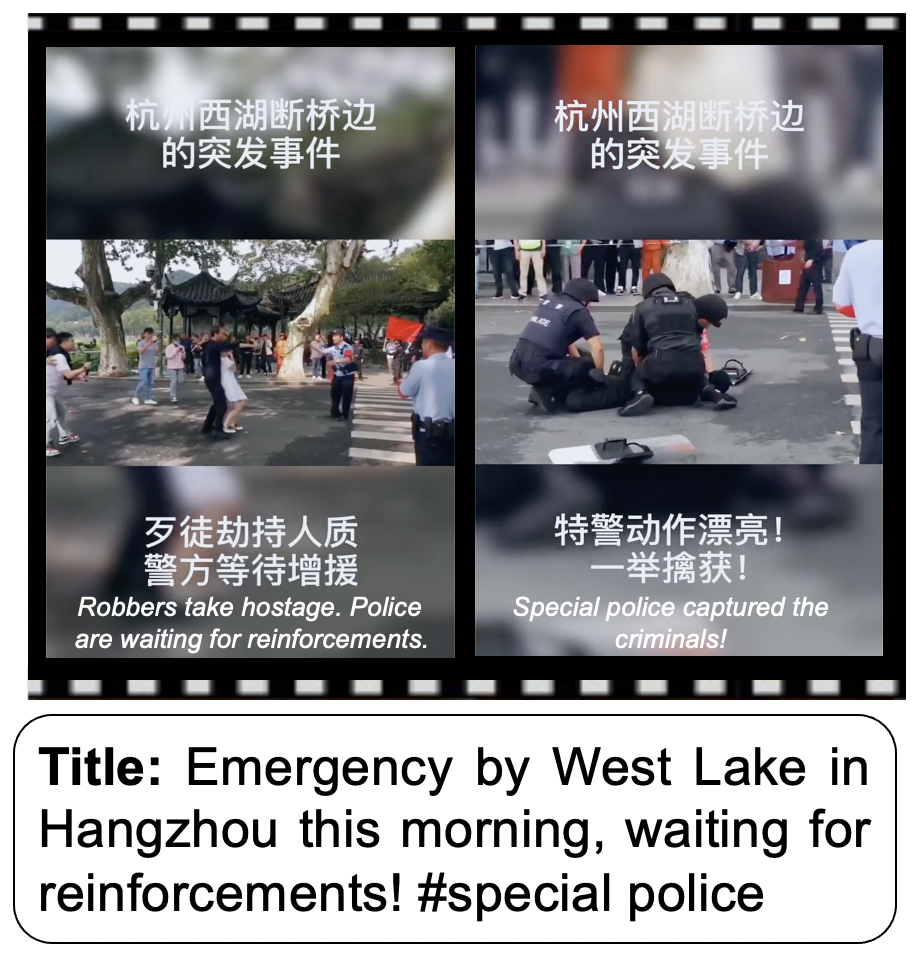}
	\label{fig:case-failed}
	}
	\caption{Two representative fake news videos in FakeSV that were detected and missed by SV-FEND respectively. }
	\label{fig:case-2}
\end{figure}

To intuitively show the ability of SV-FEND, we conduct a qualitative analysis on those successful and failed examples.
Figure~\ref{fig:case-succeed} shows a successfully detected fake news video stitched together from two videos that have similar scenes but belong to different news events.
Figure~\ref{fig:case-failed} shows a missed fake news video where the police drill was depicted as a real-life incident. 
It demonstrates the one-sidedness of videos uploaded by ordinary users and the limitations of the task of fake news detection without external information, which encourages us to further explore the combination of fake news detection and fact-checking on FakeSV.

\section{Conclusion and Potential Applications}

In this paper, we constructed the largest Chinese fake news short video dataset namely FakeSV. It contains abundant features about news content and social context, which not only provide a benchmark for different methods, but also support various researches related to fake news.  
The in-depth statistical analysis revealed the differences between fake and real news videos from multiple perspectives. 
Additionally, we provided a new multimodal baseline method and performed extensive experiments to compare the performance of different methods and modalities on FakeSV.  

In addition to fake news video detection studied in this article, the multidimensional information provided in FakeSV can be useful for other applications as follows: 
1) Detecting previously fact-checked fake news and propagation intervention. According to the phenomenon revealed in this article, fake news that has been previously fact-checked can still spread. Therefore, automatic detection of previously fact-checked fake news videos and personalized recommendations of debunked videos are indispensable. FakeSV provides debunking articles/videos corresponding to the fake news videos, which could support this task. 
2) Fake news evolution analysis. By restoring the propagation (secondary editing) chain of videos based on video similarity, we could study the lifecycles of fake news events and the intentions of fake news publishers and spreaders. 
3) User trustworthiness and susceptibility analysis. We could study the user trustworthiness by combining the published videos and the profile. By employing cross-platform analysis, we could compare users preferences and susceptibility to fake news across different platforms.

%\clearpage

\section*{Acknowledgments}
The authors would like to thank the annotators for their efforts. 
Also, we thank the anonymous reviewers for their valuable comments.
This work was supported by the National Key Research and Development Program of China (2021AAA0140203), the National Natural Science Foundation of China (62203425), the Project of Chinese Academy of Sciences (E141020). 

\section*{Ethical Considerations}
Following the literature that studied user profiles in fake news detection \cite{profile, profile2, profile3}, we collect the public information of the video publishers to study the behaviors of different user groups, \textit{i.e.,} fake and real news publishers. We have anonymized the data and clearly stated what data is being collected and how it is being used in this article. Our data is meant for academic research purposes and should not be used outside of academic research contexts.

\bibliography{aaai23.bib}

%\clearpage
\appendix

\section{Dataset Construction}

\subsubsection{Collection.} Table~\ref{tab:fcsources} lists part of the official fact-checking sites we crawled. Table~\ref{tab:re} lists some regular expressions we used to extract the key sentences from fact-checking articles. The final queries related to multiple domains are partially listed in Table~\ref{tab:queries}. 

\begin{table*}[]
\caption{Part of the official fact-checking sites.}
\centering
\scalebox{0.9}{
\begin{tabular}{p{.75\columnwidth}p{.9\columnwidth}l} 
\hline
\textbf{Source}                                                    & \textbf{Description}                                                                                                                        & \textbf{URL}                                                 \\ 
\hline
Weibo Community Management Center                         & Weibo official platform to deal with user-reported illegal posts especially misinformation.                                                   & https://service.account.weibo.com/                  \\     
Weibo Net Police & Official accounts of net polices.                                                                & ---                                                 \\
China Fact Check                                          & A platform to fact-check Chinese international news.                                                           & https://chinafactcheck.com/                         \\
Jiaozhen                                                  & A fact-checking platform operated by Tencent.                                                                                      & https://vp.fact.qq.com/                             \\
Fact Paper                                                & A global fact-checking platform operated by Pengpai News.                                                                          & https://factpaper.cn/                               \\
China Joint Internet Rumor-Busting Platform           & A fact-checking platform operated by the Cyberspace Administration of China.                                                                         & https://www.piyao.org.cn/                           \\
ScienceFacts                                              & A platform to fact-check scientific claims supported by China Association for Science and Technology.                              & https://piyao.kepuchina.cn/                         \\
Dingxiang Doctor                                          & A fact-checking platform for doctors and experts in life science.                                                                                & https://dxy.com/                                    \\
Local Rumor-Busting Platforms &  Rumor-busting platforms operated by provincial branches of the Cyberspace Administration of China. & \makecell[l]{https://py.zjol.com.cn/\\ https://py.fjsen.com/ etc.}    \\
\hline
\end{tabular}}
\label{tab:fcsources}
\end{table*}

\begin{table*}
\caption{Part of the regular expressions to extract the key sentences. }
\centering
\scalebox{0.9}{
\begin{tabular}{lp{1.5\columnwidth}} 
\hline
\textbf{Regular Expression}                        & \textbf{Sample Text}                                                                                                                                  \\ 
\hline
It is rumored on the Internet that ``(.*?)'' & \#Weibo Refutes Rumors\# It is rumored on the Internet that ``\textit{A man in Zhejiang Provenience was struck to death by lightning}"...                    \\
Rumor: (.*)                               & Rumor:~\textit{Can onions kill COVID-19 viruses}?    \\
(.*) is a rumor$\mid$(fake news)                          & \textit{It costs 100 RMB to ask elementary school students to lead the way in Zaoyang City} is a rumor!                                         \\
a (piece of)? news$\mid$video about ``(.*?)''  & Recently, a video about ``\textit{12 people committed suicide by jumping off the Caiyuanba Bridge in Chongqing City}" went viral on the Internet...  \\

\hline
\end{tabular}}
\label{tab:re}
\end{table*}

\begin{table}
\caption{Demo queries.}
\centering
\scalebox{0.9}{
\begin{tabular}{cp{.8\columnwidth}} 
\hline
\textbf{Domain} & \textbf{Sample Query} \\
\hline
Society & A cabbage sells for 38.60 RMB. \\  
Health & Cabbage is made of wax.  \\
Disaster & Shipwreck in Phuket Thailand.  \\
Culture & The United Nations declares Chinese as a global universal language. \\
Education & Elementary school students lead the police to catch illegal make-up classes. \\ 
Finance &  RMB ceases to be issued. \\
Politics & Zhejiang Province sent 100,000 ducks to Pakistan to eradicate locusts. \\
Science & Clouds can predict earthquakes in advance.  \\
Military & US military intercepts Iranian missiles.  \\
                                                                                            
\hline
\end{tabular}}
\label{tab:queries}
\end{table}

\noindent
\subsubsection{Annotation.}
Nine annotators (postgraduates) were instructed (by a Chinese guideline written by the first author and sample videos) to ensure a similar level of annotation quality across all videos.
Before the formal annotation, all annotators participated in a pilot annotation test and the Cohen’s Kappa coefficient of different annotators is 0.89, which indicates good agreement.
The videos to be labeled are assigned to annotators based on different events.
Annotators are required to classify the given video into fake, real, debunk and others. If the candidate video contains misinformation that has been debunked by the given or self-retrieved debunking articles, it will be labeled as ``fake". The candidate video can be labeled as ``real" only when the annotators have retrieved official news reports related to the reported news in this video. We do not consider videos that aren't newsworthy, don't make a verifiable claim, and don't have enough evidence to judge their authenticity. 
Furthermore, the obtained annotations were double-checked by the first and second authors, and the controversial videos were dropped. 
The overall annotation process costs about 85 hours. 

\begin{figure}[t]
\setlength{\belowcaptionskip}{-12pt}
\centering
	\includegraphics[width=.5\textwidth]{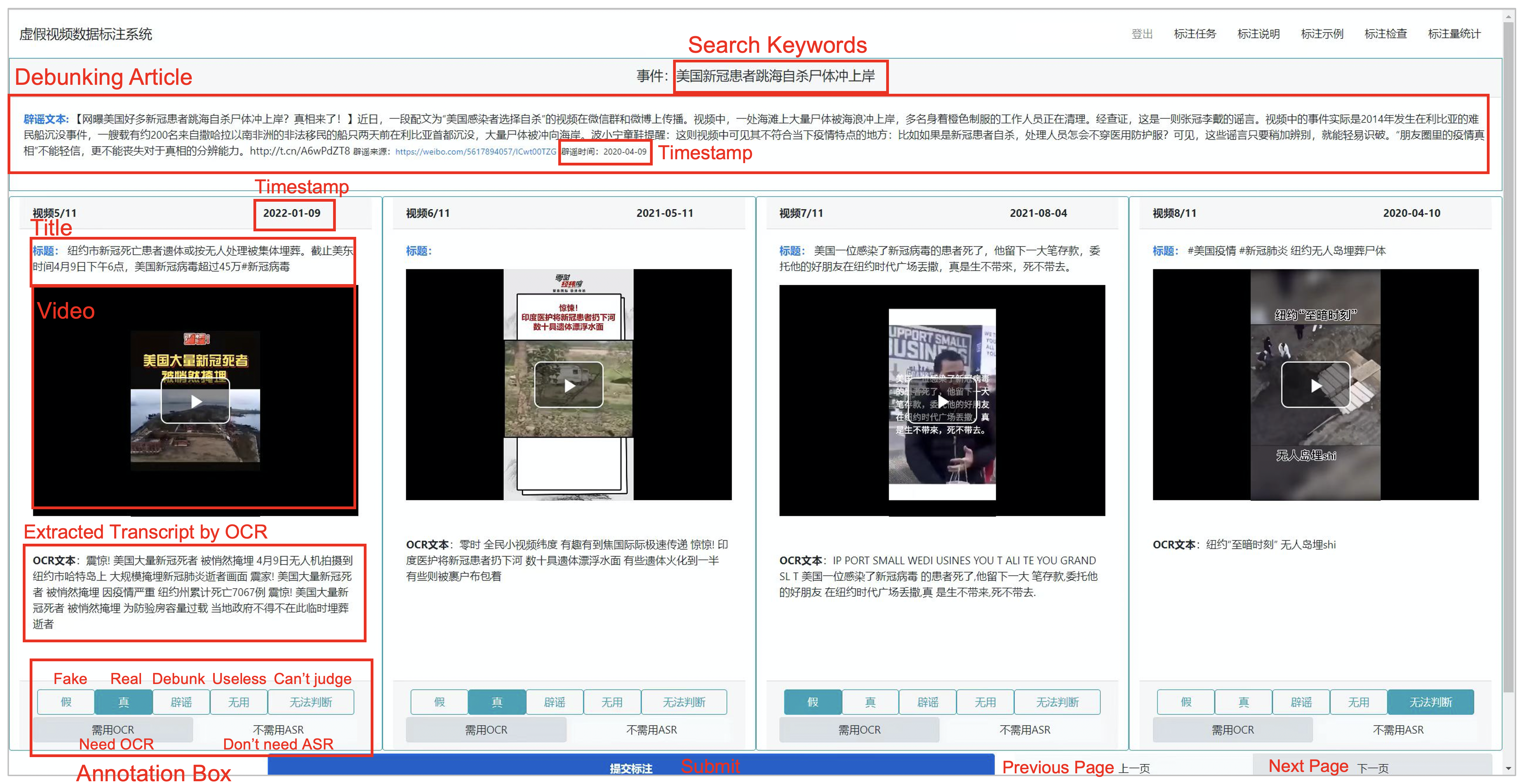}
	\caption{Annotation system.}
	\label{fig:system}
\end{figure}

Figure~\ref{fig:system} shows the developed annotation system. Considering the key evidence always exist in the text, we use the open-sourced project\footnote{https://github.com/YaoFANGUK/video-subtitle-extractor} to extract the video transcript to ease the annotation workload. 
We have also tried to recognize the speech text from audio, but the models which are pre-trained on clear actor speech datasets perform unsatisfactorily on the complex real-world audios, which are full of high-level noise, a mix of speech and music (even with lyrics), and non-standard speech (small voice and localism). 
We observe that the video transcript and speech text are highly repetitive, and thus not providing pre-extracted speech text has little impact on both annotation and detection, which has been further proved by our ex-statistics that shows 90\% videos need OCR (Optical Character Recognition) text to make a judgment while only 2\% videos need extra ASR (Automated Speech Recognition) text.

\subsubsection{Analysis on Collection Strategies. }
There are mainly two approaches to constructing datasets in existing works.
The first one is to retrieve related videos around a broad event like ``COVID-19", and then employ experienced annotators to judge the video as real or fake \cite{vavd, yt-pcancer, yt-covid, bb}. 
Due to the lack of direct evidence, this approach requires high retrieval ability and judgment of annotators. For example, \cite{yt-pcancer} and \cite{bb} hire experts to annotate the videos in specific domains. 
The second one is to search related videos around specific events which have been debunked by fact-checking sites \cite{fvc, myvc, tt}. This method reduces the demands on the annotators, who only need to judge whether the news in the video is consistent with the debunked news, and thus has the potential to obtain larger datasets with credible annotations than the first one. 
In this paper, we use the second approach to construct our dataset.
Note that the real news videos we analyzed are related to events that have been debunked instead of general real news events. We believe that it's more challenging and valuable to distinguish between fake and real news videos under the same events. Moreover, it could be further studied how a piece of real news is converted into a fake one.

\section{Data Analysis}
\label{analysis}

\subsubsection{Text. }
\begin{figure}[tbp]
	\centering
	\begin{minipage}[t]{0.23\textwidth}
		\centering
		\includegraphics[width=\textwidth]{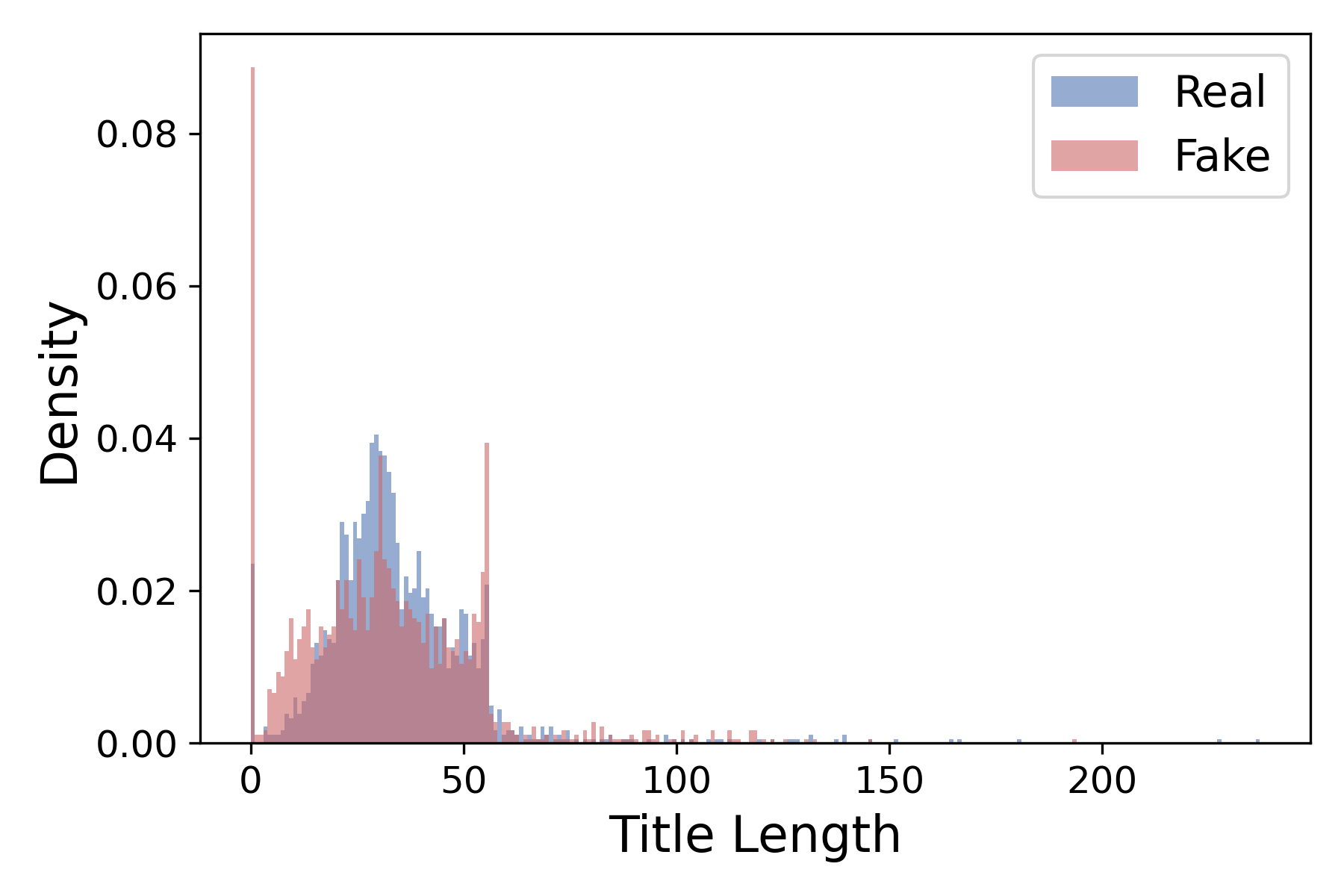}
		\caption{Length of video titles.}
		\label{fig:length}
	\end{minipage}
	\begin{minipage}[t]{0.23\textwidth}
		\centering
		\includegraphics[width=\textwidth]{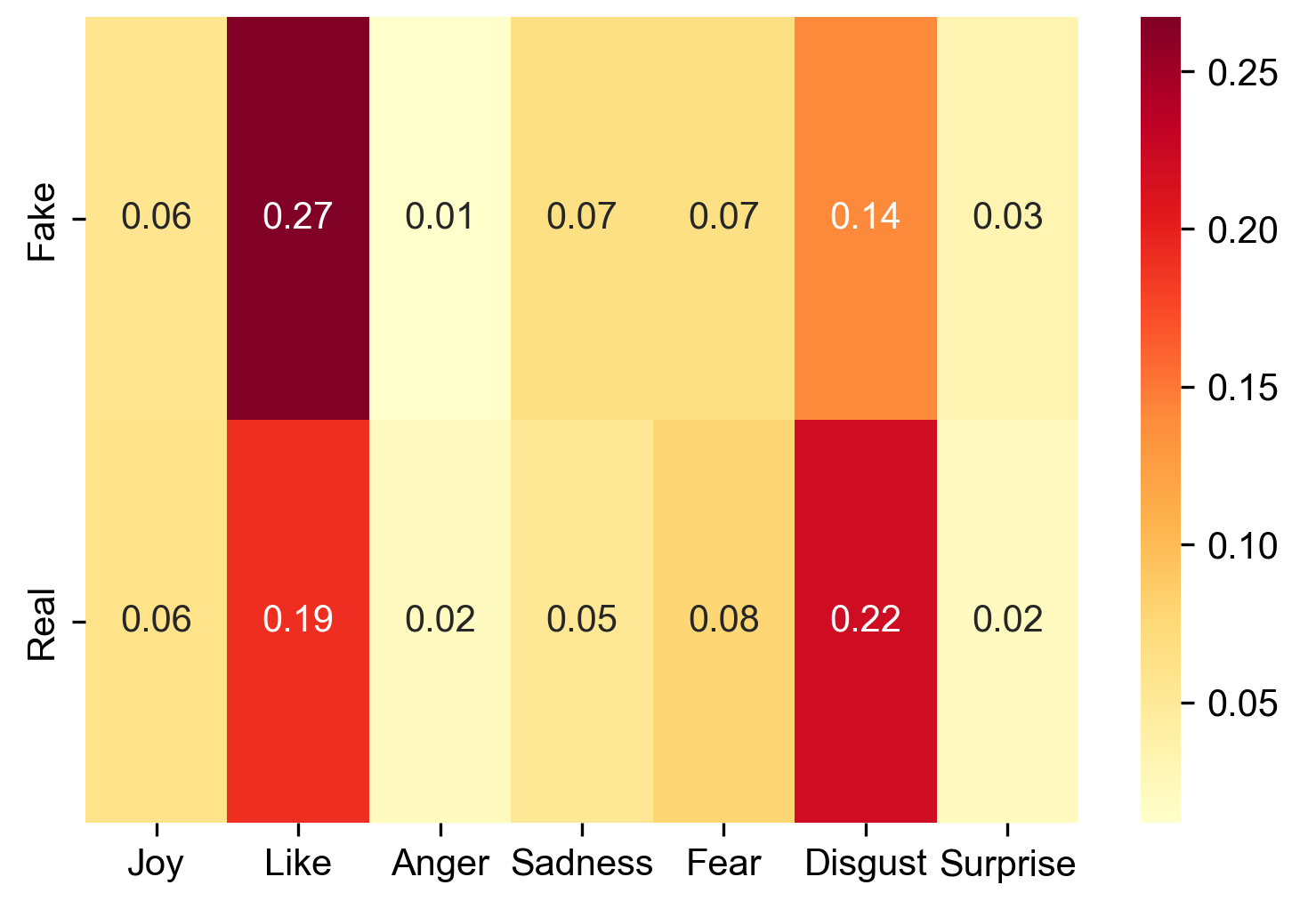}
		\caption{Emotion of video titles.}
		\label{fig:titleemo}
	\end{minipage}
	\vspace{-1em}
\end{figure}

Figure~\ref{fig:length} shows that fake news videos have shorter and more empty titles than real news. 
We also analyze the emotion of the titles. We adopt the affective lexicon ontology database \cite{emotiondb} which comprises 27,467 Chinese words belonging to seven emotion types (\ie joy, like, anger, sadness, fear, disgust, and surprise) with intensity to compute the emotion intensities. From Figure~\ref{fig:titleemo} we could see that fake news titles show more like while real news titles show more disgust.  
Considering that the video transcript is always more detailed than the title in describing the news event (avg. length: 211 v.s. 33), we also analyze the video transcript, which shows a similar distribution with the video title on text length, emotion and word usage.

\subsubsection{Spatiotemporal. }
\begin{figure}[tbp]
	\centering
	\begin{minipage}[t]{0.22\textwidth}
		\centering
		\includegraphics[width=\textwidth]{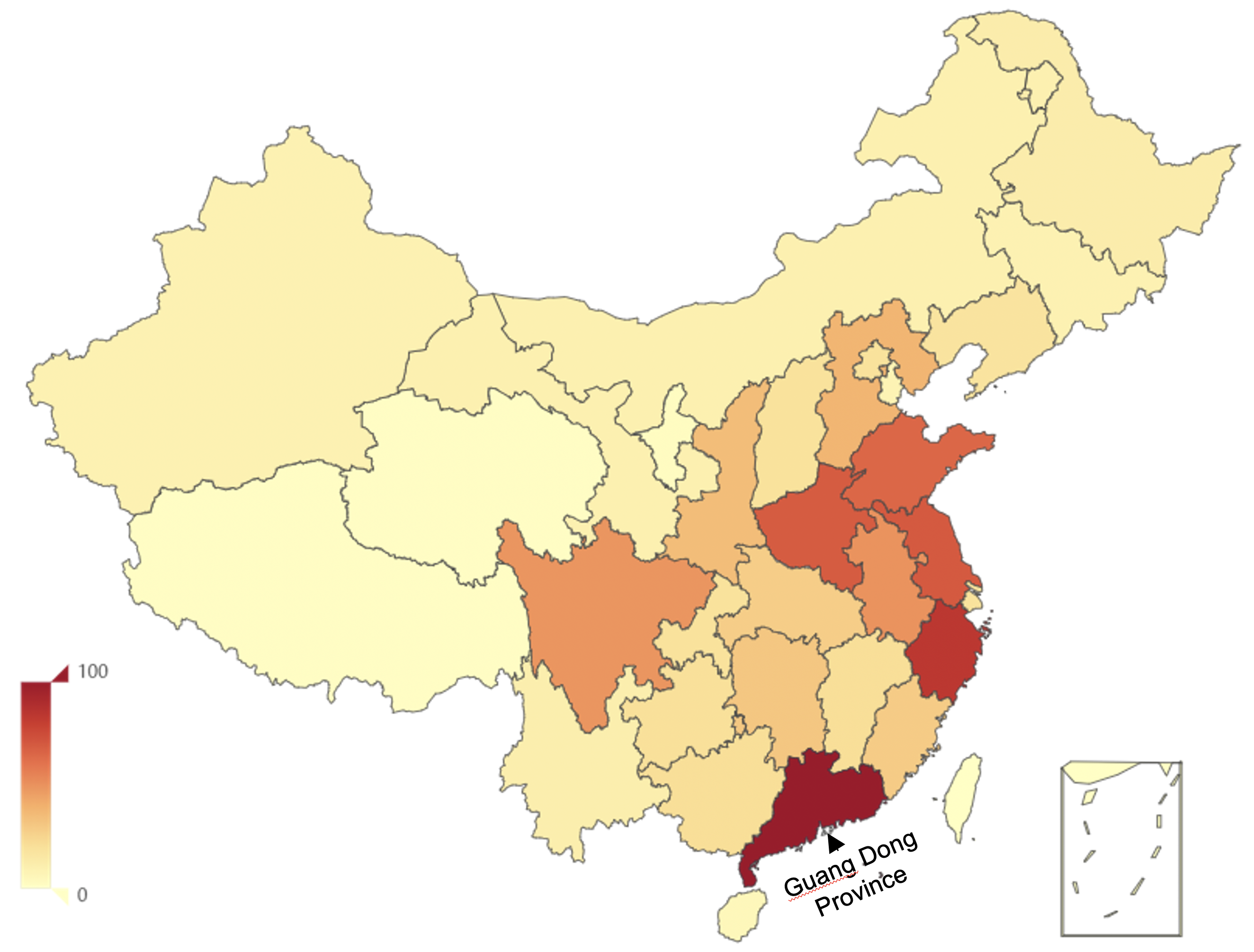}
		\caption{Spatial distribution of rumormongers.}
		\label{fig:geo}
	\end{minipage}
	\begin{minipage}[t]{0.24\textwidth}
		\centering
		\includegraphics[width=\textwidth]{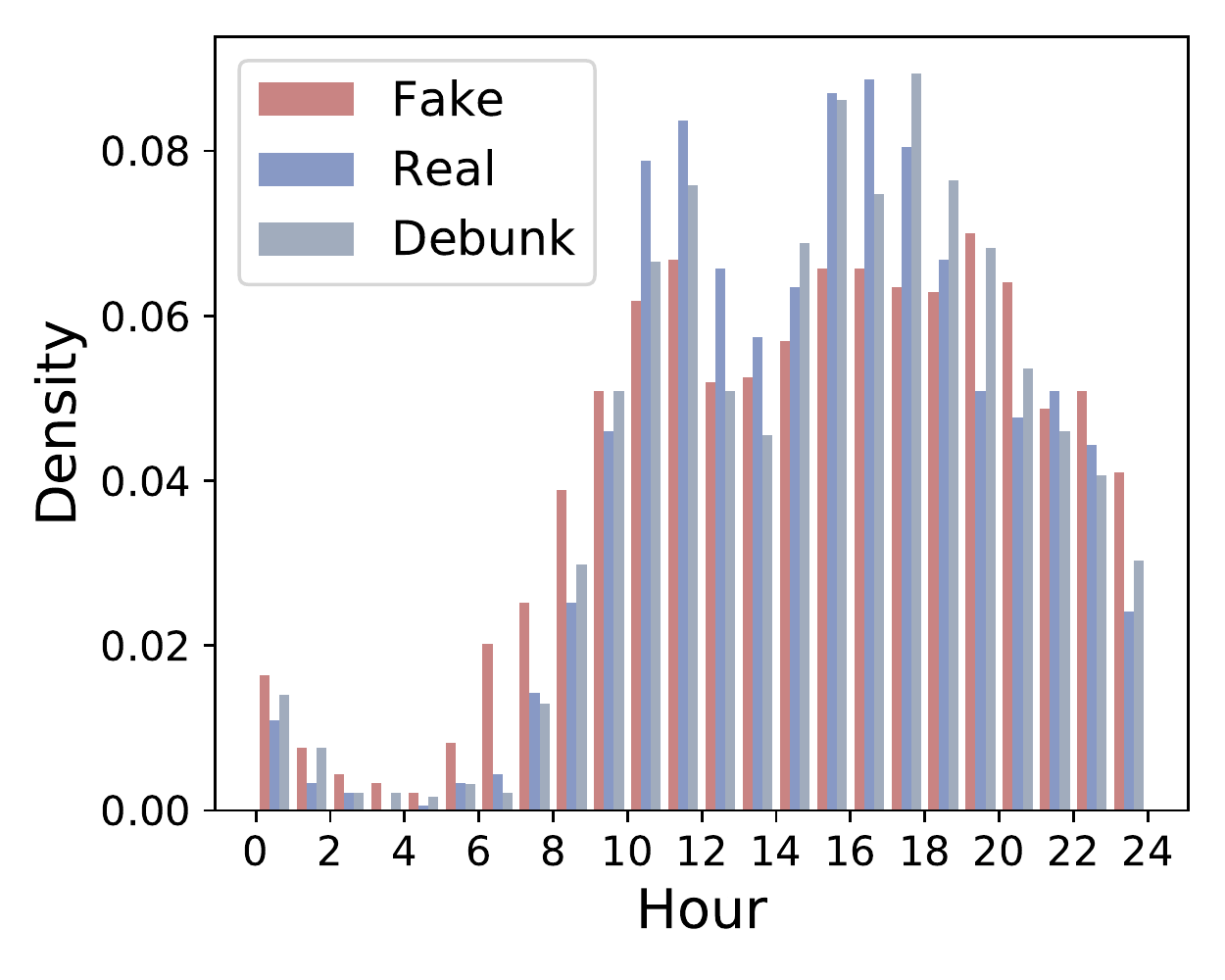}
		\caption{Distribution of published time. }
		\label{fig:hour}
	\end{minipage}
\end{figure}

We also analyze the spatial and temporal behaviors of publishers. 
Figure~\ref{fig:geo} reveals the regional differences in the geographical distribution of rumormongers based on their current IP location. We could see that Guangdong Province and many provinces along the central coast have significantly more rumormongers than other places, which would be an interesting phenomenon for developing social science studies. 
Figure~\ref{fig:hour} shows the published time of different classes of videos. We could see that the time distributions of these three classes of videos are similar, with 0:00-8:00 being the trough and 11:00/15:00 being two peaks in a day. However, fake news videos emerge more than the other in OFF hours (19:00-8:00), while real and debunking videos emerge more than fake news videos in WORKING time (10:00-18:00). 

\section{Experimental Settings}
We use the pre-trained bert-base-chinese\footnote{https://github.com/google-research/bert}, VGGish\footnote{https://github.com/harritaylor/torchvggish}, VGG19\footnote{https://pytorch.org/vision/main/models}, and C3D\footnote{https://github.com/yyuanad/Pytorch\_C3D\_Feature\_Extractor} models to extract the multimodal features. 
In the co-attention transformer block, we employ 4 heads and the hidden size is set as 128, and the number of heads is 2 for the self-attention transformer.
According to the 75\% quantile, we set the maximal number of video frames and audio frames as 83 and 50 respectively. We use zero vectors to complement and use the uniform sampling for videos that have more frames. 
Similarly, we set the maximal number of comments as 23. 
The model is trained for 30 epochs with early stopping to prevent overfitting.
We use ReLU as the non-linear activation function and use Adam \cite{adam} algorithm to optimize the loss function. The learning rate is set as $1\times10^{-4}$.
Except for feature extraction, we train our model end-to-end.

\section{Implementation of Baselines}

\begin{table}
\caption{Details of hand-crafted features.}
\centering
\scalebox{0.8}{
\begin{tabular}{p{.27\columnwidth}p{.8\columnwidth}} 
\hline
\textbf{Data} & \textbf{Features} \\
\hline
Metadata & \makecell[l]{the number of comments \\ the number of likes \\ the video duration in seconds \\ the number of videos that the publisher uploaded \\ the follower-following ratio} \\ 
\hline  
Title\&Transcript & \makecell[l]{text length \\ the number of words \\ contains question/exclamation mark (Boolean) \\ contains 1st/3rd person pronoun  (Boolean)\\ the number of positive/negative sentiment words\\ has ``:'' symbol (Boolean) \\ the number of question/exclamation marks\\ has clickbait phrase (Boolean)\\ sentiment polarity \\ the number of modal particles \\ the number of personal pronouns \\ tf-idf \\ Ngrams \\ LIWC }  \\
\hline  
Comments & \makecell[l]{comments fakeness ratio \\
comments inappropriateness ratio (swear words) \\
comments conversation ratio (at least one reply)\\
top 100 comments tf-idf \\
top three popular comments: sentiment polarity, \\ the number of modal particles, the number of \\ personal pronouns and text length \\
} \\
\hline
\end{tabular}}
\label{tab:hcfeas}
\end{table}

The implementation details of single modality baselines are as follows:
\begin{itemize}
    \item \textbf{hc feas + SVM:} We use the implementation of LinearSVC available in scikit-learn\footnote{https://scikit-learn.org/}. We normalize each feature using L2 norm before feeding it into the classifier. We adapt the hand-crafted features to our dataset, and list them in Table~\ref{tab:hcfeas}. When experimenting with methods only based on comments, data without comments are filtered out.
    \item \textbf{Text-CNN: } We implement Text-CNN which has filter windows of 3, 4, 5 with 14 kernels each. 
    \item \textbf{BERT: } We use the pre-trained BERT without fine-tuning. We truncate the sequences to the maximum length of 512 and 256 when dealing with the title\&transcript and comments, respectively.
    \item \textbf{VGG19+Att:} We extract features from the first fully-connected layer of VGG19 model pre-trained on the ImageNet dataset.
    \item \textbf{Faster R-CNN+Att:} We use the implementation\footnote{https://github.com/chuhaojin/BriVL-BUA-applications/tree/master/bbox\_extractor} to extract the bounding boxes and use VGG19 to extract the object features. 
    \item \textbf{C3D+Att:} We extract features from the first fully-connected layer of C3D model pre-trained on the Sport1M dataset.
    \item \textbf{emo feas:} We use the open-sourced project OpenSmile\footnote{https://audeering.github.io/opensmile/get\-started.html\#obtaining\-and\-installing\-opensmile} to extract three types of audio emotion features, \textit{i.e., Emo\_IS09, Emobase and Emo\_large}, and show the best results.
    \item \textbf{VGGish:} We freeze the convolutional layers and fine-tune the rest layers. 
\end{itemize}

We implement multimodal baseline methods based on descriptions in their original papers due to the unavailability of the original codes. All of these codes will be released to promote fair comparisons. 
\begin{itemize}
\item \textbf{\cite{yt-pcancer}:} 
 Considering that the extracted OCR text is more accurate than the ASR text and the high similarity of them, we extract the linguistic features of the OCR text instead of ASR. 
 We use the implementation available in LIWC lexicon\footnote{https://cliwceg.weebly.com/} to extract the Ngrams and psycholinguistic features. We use the emo\_base features to replace the emo\_IS09 features and then use the MIC (Maximal Information Coefficient) to select the top 10 acoustic features for obtaining better performance. 
\item \textbf{\cite{yt-covid}:} 
We set the word frequency threshold as 5 when extracting the tf-idf features. The ratio of conspiracy comments was ignored due to lacking the annotated data.
\item \textbf{\cite{myvc}:} We set the maximal number of video frames and comments and  as 83 and 23, respectively. 
\item \textbf{\cite{tt}:} We use the public API\footnote{https://console.cloud.tencent.com/asr} to extract the ASR text and use the librosa library\footnote{https://librosa.org/} to extract the MFCC feature of each speech word. The maximum of the sequence length is set as 100 and 500 in CVRL and ASRL modules, respectively.  
\end{itemize}

\end{document}